\documentclass[preprint,12pt]{elsarticle} 
\usepackage[utf8]{inputenc}
\usepackage{amsmath}
\usepackage{amssymb}
\usepackage{hyperref}
\usepackage{graphicx}
\usepackage{float}
\usepackage{array}
\usepackage{multirow}
\usepackage{hhline}
\usepackage{caption}
\usepackage{subcaption}
\usepackage{tikz}
\usetikzlibrary{shapes}
\usetikzlibrary {calc,intersections,through}
\journal{Journal of Theoretical Biology}

\cortext[cor1]{Corresponding Author. Email at jesse.zhou@research.uwa.edu.au}
\date{}

\begin{document}
\begin{frontmatter}
\title{A selfish herd with a target}  
\author[add:UWA]{Jesse Zhou \corref{cor1}}
\author[add:UWA]{Shannon Dee Algar}
\author[add:UWA]{Thomas Stemler}
\address[add:UWA]{The University of Western Australia, Department of Mathematics and Statistics, Crawley, 6009, Australia}

\begin{abstract}
    One of the most striking phenomena in biological systems is the tendency for biological agents to spatially aggregate, and subsequently display further collective behaviours such as rotational motion. One prominent explanation for why agents tend to aggregate is known as the selfish herd hypothesis (SHH). The SHH proposes that each agent has a ``domain of danger" whose area is proportional to the risk of predation. The SHH proposes that aggregation occurs as a result of agents seeking to minimise the area of their domain. Subsequent attempts to model the SHH have had varying success in displaying aggregation, and have mostly been unable to exhibit further collective behaviours, like aligned motion or milling. 
    Here, we introduce a model that seeks to generalise the principles of previous SHH models, by allowing agents to aim for domains of a specific (possibly non-minimal) area or a range of areas and study the resulting collective dynamics. Moreover, the model incorporates the lack of information that biological agents have by limiting the range of movement and vision of the agents. The model shows that the possibility of further collective motion is heavily dependent on the domain area the agents aim for - with several distinct phases of collective behaviour. 
\end{abstract}
\end{frontmatter}
\section{Introduction}
One prevalent and striking phenomena in biological systems is the emergence of collective order as the number of individual particles or organisms in the system grows large.\cite{Attanasi, ALGAR201982, COUZIN20021} A system begins to exhibit collective order when the behaviours and states of the individuals synchronise, giving the impression of a behaviour or property of the collective as a whole\cite{Attanasi}. A common collective behaviour is the formation of aggregates\cite{HAMILTON1971295, Wood2007-np, Parrish1999-cf} : individual fish and birds for example, will often be observed grouping together to form shoals in open waters and flocks respectively\cite{HAMILTON1971295, Baerends1950, williams_1964}. Upon the formation of an aggregate, the fish or birds in the aggregate can then display further collective order, such as travelling with uni-directional movement and forming complex structures such as vortices or v-shaped formations\cite{Attanasi, ALGAR201982, COUZIN20021}. The initial formation of the aggregate and subsequent collective behaviour is defined as flocking, and understanding how and why flocking occurs remains a central question in the fields of biology and evolution.

One particularly influential explanation of flocking is the Selfish Herd Hypothesis (SHH), which proposes that flock formation occurs as a result of anti-predation responses. First put forward by Hamilton in 1970, the SHH is based on the premise that flocking agents have no information on the location of potential predators; therefore, the risk of predation for a given agent is proportional to the area of the domain closer to that agent than any other - Hamilton defined such an area as the \textit{domain of danger} (DOD). Mathematically, the domain of danger is known as the Voronoi cell\cite{voronoibook}, and in two dimensions, is a convex polygon. Hamilton postulated that flock formation occurs as a consequence of the agents seeking to minimise their predation risk by minimising their DODs. Subsequent work sought to confirm the SHH by evaluating the effectiveness of various movement strategies in minimising the DODs and flock formation. However, in all previous approaches, once the initial aggregate was formed, no further collective order would be observed as the group would remain in a stationary state as the flock would reach a compact state where alternative positions for agents were either worse off or were unreachable\cite{MORTON199473, JAMES2004107, VISCIDO2002183}. Previous models are therefore inadequate for providing an explanation of further collective behaviour (flocking) that could be exhibited by a system once the aggregate is established, such as rotational or uni-directional collective movement.

A recent paper by Monter et al.\cite{MONTER2023111433} however, overcame this problem by introducing two key features to their model. Firstly, a reinforcement learning scheme was used to develop the movement decision policy used by the agents during simulations. Secondly, many animals in nature do not have all-encompassing vision - the field of vision for different species extends to different angles, and \cite{MONTER2023111433} sought to introduce this fact by limiting the model agents' neighbour detection to 90 degrees either side of the direction the agent was facing. Inspired by \cite{MONTER2023111433}, we investigate a model that attempts to drive further collective motion and rotational order after flock formation. We do so by generalising the movement optimisation scheme of \cite{ALGAR201982}, with agents aiming for a DOD of a specific area, rather than just a minimal area. In addition, we use a model that restricts agents' information about the environment behind them, and therefore only allow them to compute an incomplete or bounded DOD. We introduce this model, along with an alternative method of calculating a circle bounded DOD, in further detail in the next section. Simulation parameters and measures of the collective order are then introduced in section 3. Finally, in section 4, we show that such a model is able to drive continuous motion under appropriate parameter sets, and can also exhibit further characteristics and phases of collective order. 

\section{Methodology}
\subsection{Voronoi cells and the domain of danger}
\begin{figure}[H]
        \centering
        \begin{subfigure}[t]{0.3\textwidth}
         \centering
         \includegraphics[height=\textwidth]{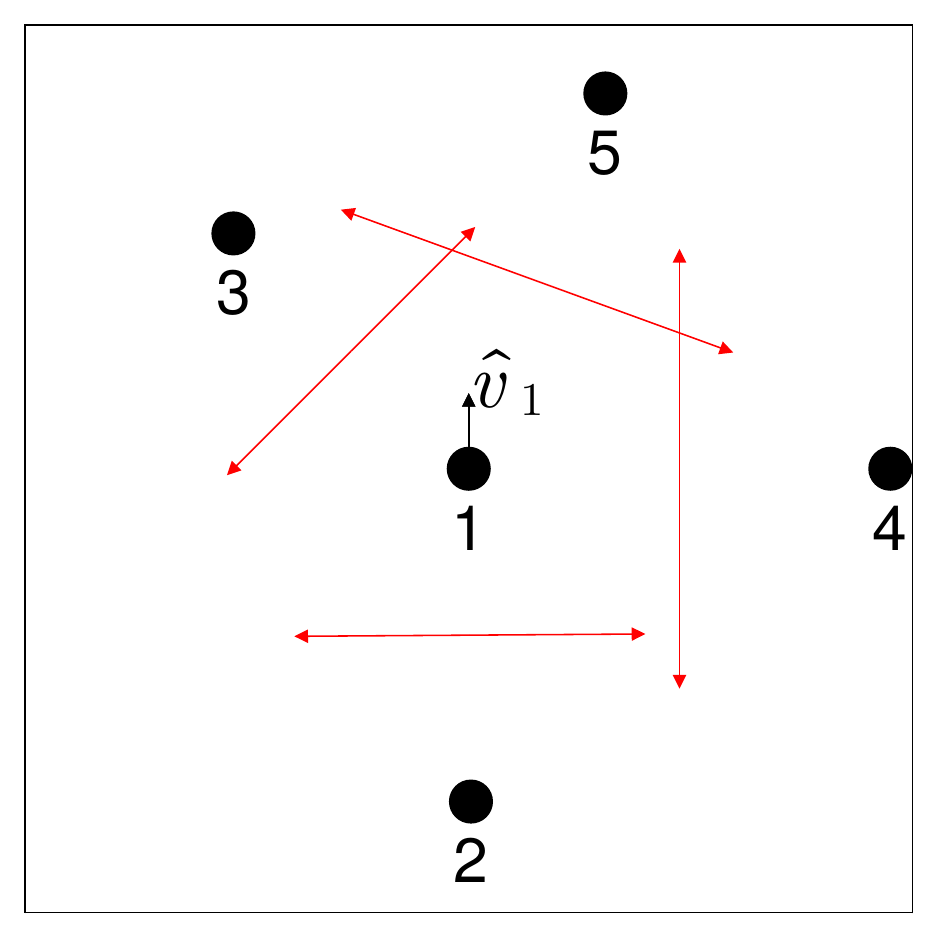}
         \label{fig:y equals x}
         \caption{}
     \end{subfigure}
     \hspace{1cm}
     \begin{subfigure}[t]{0.3\textwidth}
         \centering
         \includegraphics[height=\textwidth]{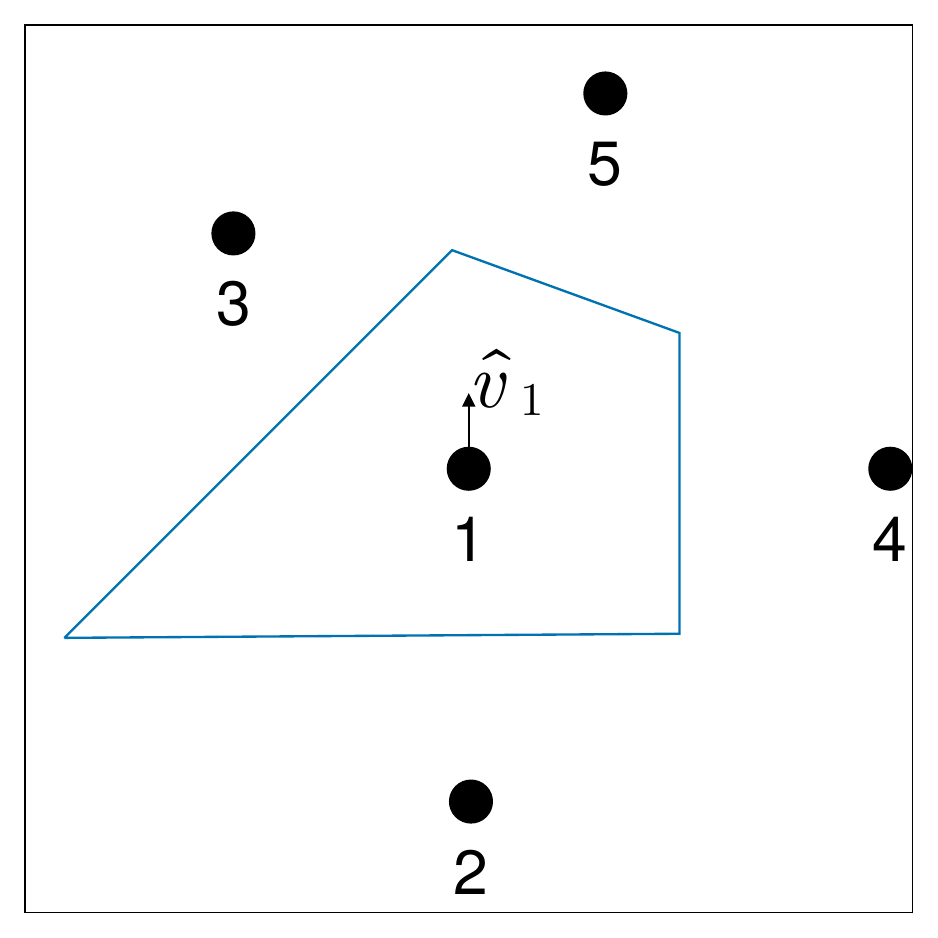}
         \caption{}
         \label{fig:half_plane_voronoi}
     \end{subfigure}
     \caption{An illustration of the half plane approach to Voronoi cell calculations.  In figure (a), an illustration of the ``fences" (red lines) between an agent $1$ and its neighbours - agents $2, 3, 4, 5$. Note that the arrow labeled as $\hat{v}_1$ indicates the direction the agent is facing. Figure (b) gives an illustration of the Voronoi cell of the agent $1$ from (a). We note that the Voronoi cell can be derived from the intersection of its ``fence'' - this idea serves as the basis for the half plane intersect algorithm.}
     \label{fig:half_plane_approach}
\end{figure}

The concept of the Voronoi cell, or domain of danger as termed by Hamilton, is critical to his theory of flock formation. In this section, we give the formal definition of the Voronoi cell and outline an  approach to calculating the circle bounded Voronoi cell of an agent that can run in $n\log(n)$ time, where $n$ is the number of agents in our model. Mathematically, the domain of danger  of an agent $i$ - which we denote as $A_i$ - residing in an $d$-dimensional space is given by the points $\Vec{x}$ such that
    \begin{align}
        A_i = \{\Vec{x}\in \mathbb{R}^d: ||\Vec{x}- \Vec{r}_i|| \leq || \Vec{x} - \Vec{r}_j|| \quad \forall j\neq i\},
    \end{align}
where $\Vec{r}_i, \Vec{r}_j \in \mathbb{R}^d$ are the positions of agent $i$ and agent $j$, and $||\cdot ||$ is the typical Euclidean norm for $n$-dimensional space. In two dimensions, a Voronoi cell is a convex polygon. One of the most well-known algorithms for calculating a two dimensional Voronoi cell is known as Fortune's algorithm\cite{Fortunes}, but a method based on half plane intersections is also possible\cite{Wuhalfplane}. In this paper, we will use the half plane based approach, as this approach is naturally conducive to representing the forward limited vision of agents. To calculate the domain of an agent $i$ in the half plane approach, we make the following observation. Suppose that we draw a line from agent $i$ to one of its neighbours, $j$. We then take the perpendicular bisector of the line, which forms a “fence” half way between agent $i$ and $j$. The set of all points closer to $i$ than $j$ is the set of all points on the same side of the fence as $i$ - formally, it is the half plane which contains $i$. Then, making this observation for all neighbouring agents $j$, it is evident that the Voronoi cell of agent $i$ is the intersection of all such half planes with its neighbours. This idea is illustrated in Figure \ref{fig:half_plane_approach} and is the basis of a Voronoi cell calculation algorithm called the half-plane intersection algorithm\cite{Wuhalfplane}, which runs in $O(n(\log(n))$ time, where $n$ is the number of agents in the model. As in \cite{JAMES2004107} and \cite{ALGAR201982} however, our model will bound the area of each DOD by a circle of radius $\rho$. In doing so, we will have to make a minor adjustment in the typical half-plane intersection algorithm to account for the bounding circle. The use of a circle bounded Voronoi cell is illustrated in Figure \ref{fig:circled}.

\subsubsection{Bounded DODs and angles of vision}
\begin{figure}[H]
     \begin{subfigure}[t]{0.3\textwidth}
         \centering
         \includegraphics[height=\textwidth]{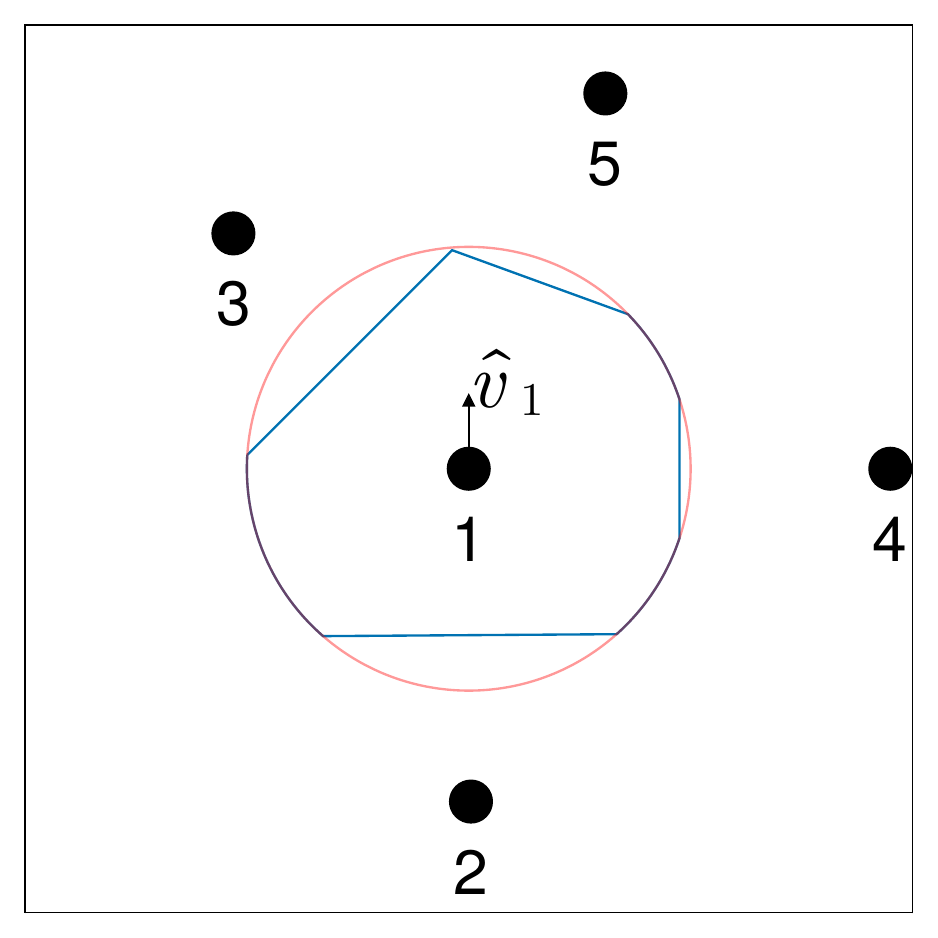}
         \caption{}
         \label{fig:circled}
     \end{subfigure}
     \hfill
     \begin{subfigure}[t]{0.3\textwidth}
         \centering
         \includegraphics[height=\textwidth]{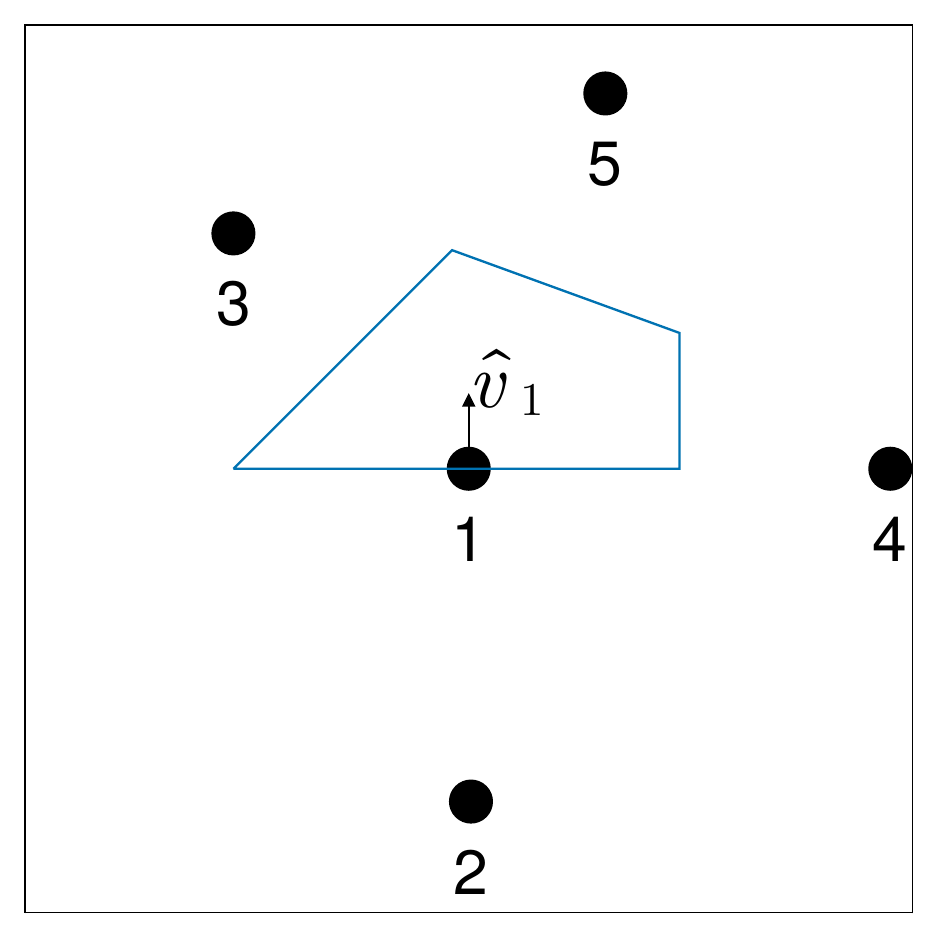}
         \caption{}
         \label{fig:forward_only}
     \end{subfigure}
     \hfill
     \begin{subfigure}[t]{0.3\textwidth}
         \centering
         \includegraphics[height=\textwidth]{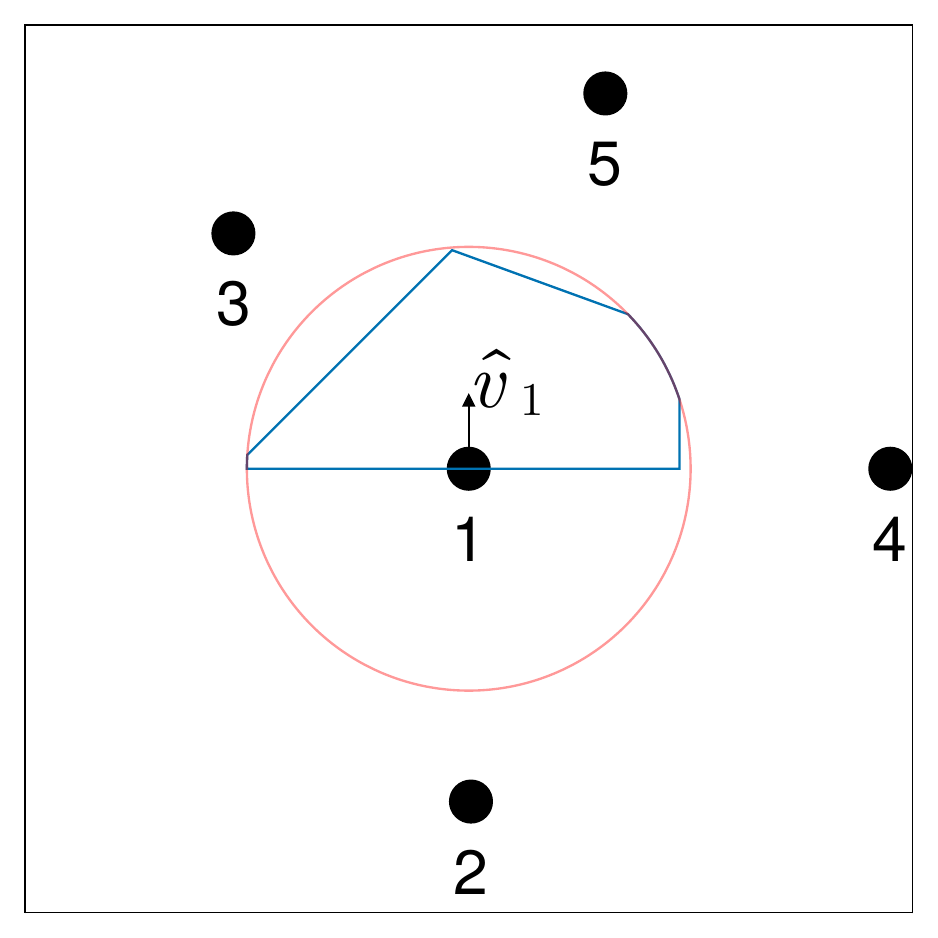}
         \caption{}
         \label{fig:forward}
     \end{subfigure}
   \caption{Illustrations of the different bounding techniques for Voronoi cell calculations using the same agent configuration as in figure \ref{fig:half_plane_approach}. Figure (a)  gives the same Voronoi cell as in figure \ref{fig:half_plane_voronoi}, but now bounded by the red circle. Figure (b) gives the forward bounded version of the Voronoi cell calculated in figure \ref{fig:half_plane_voronoi}. Finally, figure (c) represents the DOD agent 1 calculates in our model, using a forward bounded DOD also bounded by a circle.}
    \end{figure}

One major difference between our model and previous models of Hamilton's SHH is that previous models assumed that agents have a complete view of the neighbourhood around them, and would therefore be able to calculate their full DOD. However, in reality, the field of view for agents varies greatly - while certain birds have been cited as having an almost complete $360$ degree field of view \cite{MARTIN200825}, with humans for example, the field of view is commonly cited to be in the range of $180-220$ degrees\cite{doi:10.1177/2041669520913052}.  In order to represent the limited view of agents then, when an agent tries to calculate a Voronoi cell, our model restricts the Voronoi cell calculation such that only the section of the Voronoi cell ``in front" of the agent, given their current position and heading, is accounted for. We call DODs calculated with this restriction \textit{forward-bounded DODs}, and this method is illustrated in figure \ref{fig:forward_only}. Ultimately, since we also use this technique in conjunction with a bounding circle, the agents calculate cells like that illustrated in figure \ref{fig:forward}.

The algorithm we use to calculate the forward and circle bounded Voronoi cell of an agent is detailed and illustrated in the appendix. The code for the algorithm can be found in our Github repository \cite{Zhou_Selfish_Herd_and_2024}.

\subsection{Movement scheme}
\begin{figure}[H]
        \centering
        \begin{subfigure}[t]{0.45\textwidth}
         \centering
         \includegraphics[width=\textwidth]{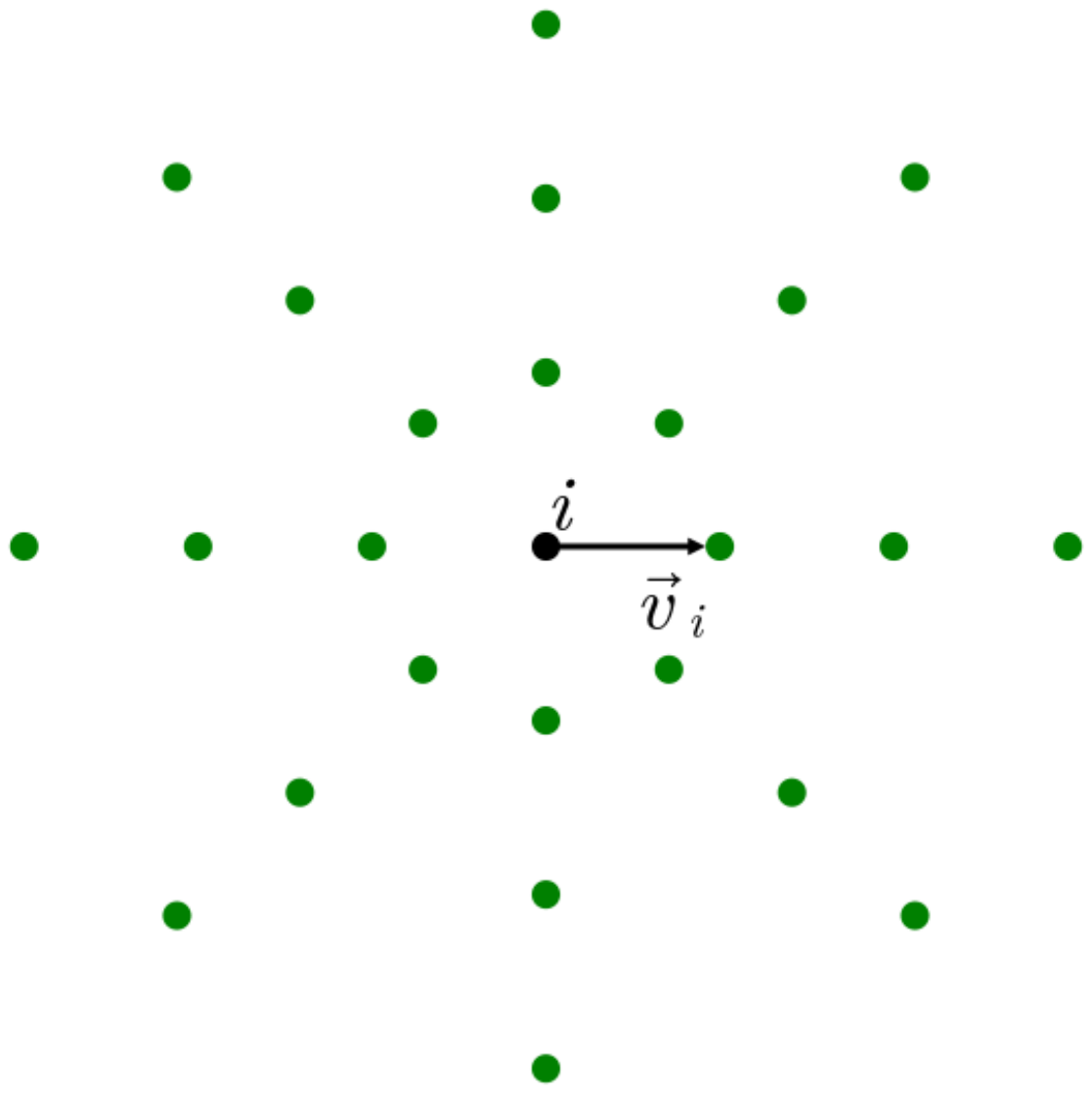}
         \caption{}
         \label{fig:sampling_algar}
     \end{subfigure}
     \hfill
     \begin{subfigure}[t]{0.45\textwidth}
         \centering
         \includegraphics[width=\textwidth]{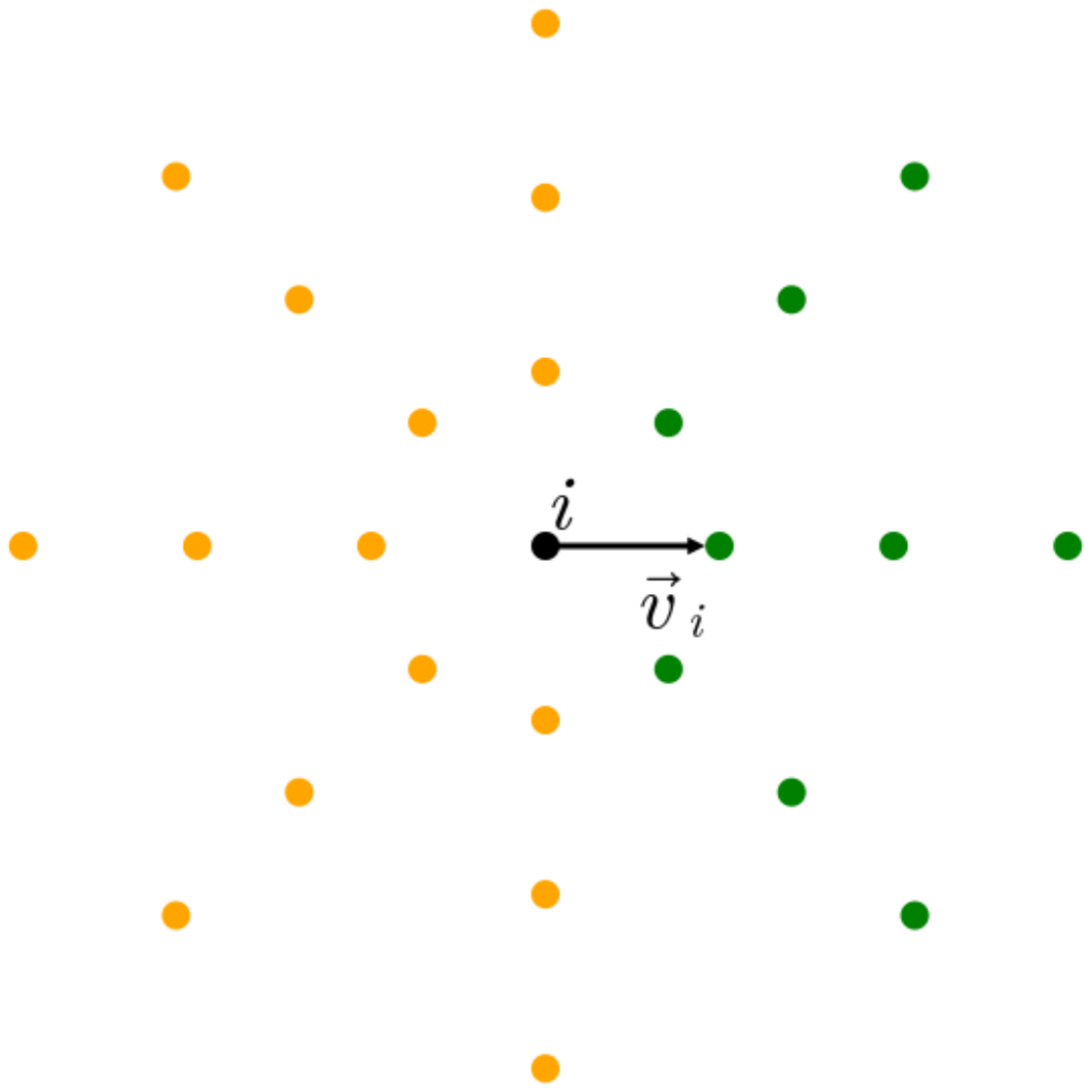}
         \caption{}
         \label{fig:sampling_stdod}
     \end{subfigure}
     \caption{(a) An illustration of the angular sampling scheme in \cite{ALGAR201982} - the agent samples $q = 8$ equally spaced angular coordinates and $m = 3$ radial coordinates. (b) An illustration of the limited angular sampling scheme used in our model: we still sample $m = 3$ radial coordinates, but of the original $q = 8$ angular coordinates, we only sample 3 angular coordinates to the front of the agent ($q' = 1$), with the agents' current velocity indicated by $\Vec{v}_i$.}
    \label{fig:sample_ill}
    \end{figure}

Given its success in demonstrating flock formation, we will use a movement scheme based on the model proposed in \cite{ALGAR201982}. As in \cite{ALGAR201982}, each particle or agent is represented as a circular disc with a radius defined to be one \textit{body length} (BL), which will be used as the unit of length throughout this document. At each time step, each agent samples positions along the direction of $\theta_0$, the angular coordinate of its current velocity, as well as the angles
    \begin{align}
        \theta_0 \pm 1\times \frac{2\pi}{q}, \theta_0 \pm 2\times \frac{2\pi}{q},..., \theta_0 \pm q'\times \frac{2\pi}{q},
    \end{align}
where $q, q'$ are integers such that $q' < q$. In other words, each agent samples positions along the direction of its current heading, as well as the $q'$ closest directions to either side of $\theta_0$, with a spacing of $\delta_q = \frac{2\pi}{q}$ between adjacent angles. In each of these directions, the agent samples positions a distance of $1, 2, 3..., m$ BL away from its current position. We illustrate this sampling method in \ref{fig:sample_ill}

To enforce volume exclusion (so that two agents are not occupying the same space), if a probed position is of distance less than two $BL$ from the position of another agent at that time step, then that position is not considered for optimisation. Moreover, if an agent detects that moving 1 BL in a certain direction will cause a collision (i.e, violate volume exclusion), the agent will not sample any positions along that angular coordinate. 

In contrast to previous models of the SHH, we seek to generalise the principles behind the SHH such that, instead of always seeking a position with the minimal DOD, agents seek positions with a DOD as close to a target area that we set as a model parameter. We define the DOD that the agents seek as the \textit{target DOD} or \textit{tDOD}. Thus, when sampling positions, agents will determine the optimal position by how close the DOD in such a position is compared to the target DOD. The agent then moves one body length in the direction of the sampled position that optimises the domain area. Critically however, if none of the sampled positions have a potential DOD closer to the target DOD than the agent’s current position, the agent remains stationary but makes a minimal turn on the spot to either their left or right. In other words, they set the direction of their velocity randomly to either $\theta_0 - \delta_q$ or $\theta_0 + \delta_q$, but their speed will be 0. Allowing agents to turn when they are unable to find better positions under the limited angle sampling scheme is necessary for flock formation  - otherwise, agents on the convex hull with initial velocities facing outwards will only see positions in the open domain with the largest possible DODs. For the majority of tDODs then, such a scheme would result in poor flock formation as the agents on the convex hull would remain unmoving, often leading to other agents also being attracted to them, forming a number of small sub-flocks. 
    \begin{figure}[H]
        \centering
        \includegraphics[width=0.5\textwidth]{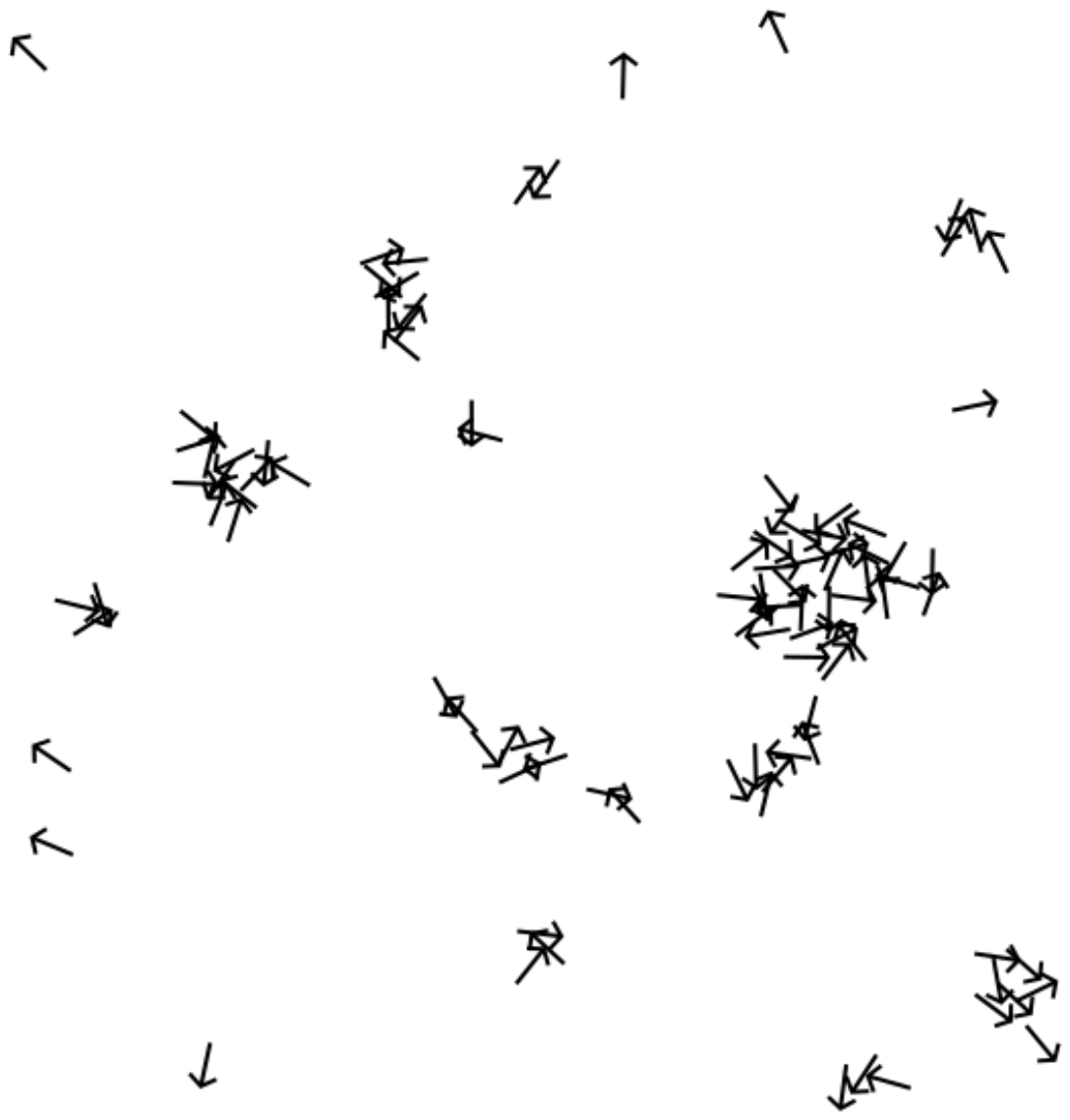}
        \caption{An illustration of the flock formed when agents cannot perform a stationary turn. A video of how the flock behaves without turning is also provided in supplementary video 1. We see that agents initialised on the edge of the group often remain there due to being unable to see better positions. }
        \label{fig:crystal_state_sample}
    \end{figure}

The position of each agent is updated at each time step, with discrete time increments of size $\Delta t$. The model uses the Euler method to determine the position of the agent at each successive time step, given by
    \begin{align}
        \Vec{r}_i(t + \Delta t) = \Vec{r}_i(t) + \hat{v}_i(t) s_i(t)\Delta t 
    \end{align}
where $\Vec{r}(t)$ is the position of the agent at time $t$ and $\hat{v}_i(t)$ is the unit vector for the direction of travel determined by the movement rule previously defined. $s_i(t)$ is the speed of the agent at time $t$ and is equal to 0 if the agent is unable to find a better position than their current position, and 1 if the agent is able to find a more optimal position. 


We also note that in the following sections, our choice of the tDODs to explore is motivated by a result known as Thue's theorem\cite{Chang2010ASP}. Thue's theorem  postulates that the densest possible packing configuration for equally sized circles in two dimensional space is a hexagonal configuration.
From this, it can be derived that the volume occupied by each ball of radius $1$ in the densest packing configuration is $\sqrt{12}$. Thus, in our measurements of area and in our choices for the tDODs that we explore in subsequent sections, we do so with reference to the minimal area of $\sqrt{12}$.



\section{Simulation and Measures}
The model was simulated with 100 agents randomly initialised on a $100\times100$ grid. At every time step, agents sampled radial coordinates from 1, 2,..100BL away from their current position.
The agents also generated $q = 8$ equally spaced angular coordinates, but would only sample $q’ = 1$ coordinates to either side of their current velocity for potential new positions. We found that the total of three angular coordinates was enough to capture the collective behaviours we aimed for. We use a vision radius of $\rho = 100$ and tested model dynamics sweeping through a range of target Voronoi areas for the agents from $0.0$ to the maximum of $\pi\times \rho^2$. For each target DOD, we run 20 simulations with 5000 time steps in each simulation. We ran each simulation in a closed, non-periodic domain typically of dimensions $1000\times1000$. For larger tDODs however, we would increase the domain size to $5000\times 5000$. The two main collective behaviours which we are interested in are rotation or milling, and continued motion. To track rotational motion, we use the rotational order parameter as defined by \cite{MONTER2023111433} and \cite{COUZIN20021}, given by 
    \begin{align}
        \Phi_\mathrm{R} = \frac{1}{n}\sum_i|\hat{r}_i\times \hat{v}_i|,
    \end{align}
where $\hat{r}_i, \hat{v}_i$ represent the unit vectors in the direction of the position of the $i$-th agent relative to the group center, and the velocity of the $i$-th agent respectively. The rotational order parameter can therefore have a value between 0 and 1, and is a measure of the degree of rotation of the agents around the center of the group. To track continued motion, we use the mean speed parameter given by
    \begin{align}
        \Phi_S = \frac{1}{n}\sum_{i = 1}^N s_i,
    \end{align}
where $s_i$ is the speed of the $i$-th agent. 

\section{Results and Discussion}
Videos illustrating the flock behaviours in this section can be found as supplementary videos in the project Github repository \cite{Zhou_Selfish_Herd_and_2024} under the folder ``Records/Thesis".
\subsection{Results}
\begin{figure}[H]
        \begin{subfigure}[t]{0.5\textwidth}
            \centering
            \includegraphics[width=\textwidth]{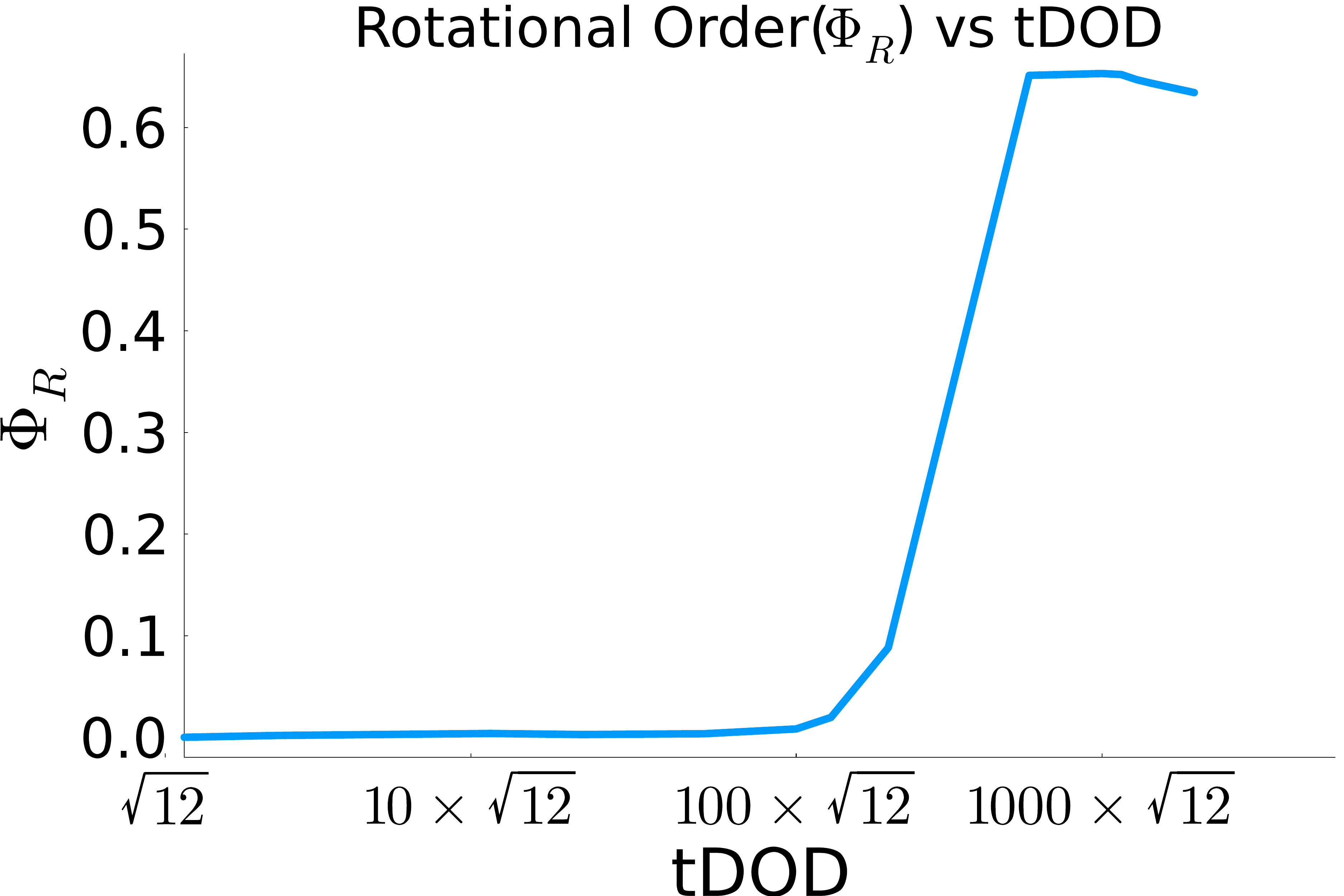}
            \caption{}
            \label{fig:rot_o_alt_vs_tdod}
        \end{subfigure}
        \begin{subfigure}[t]{0.5\textwidth}
            \centering
            \includegraphics[width=\textwidth]{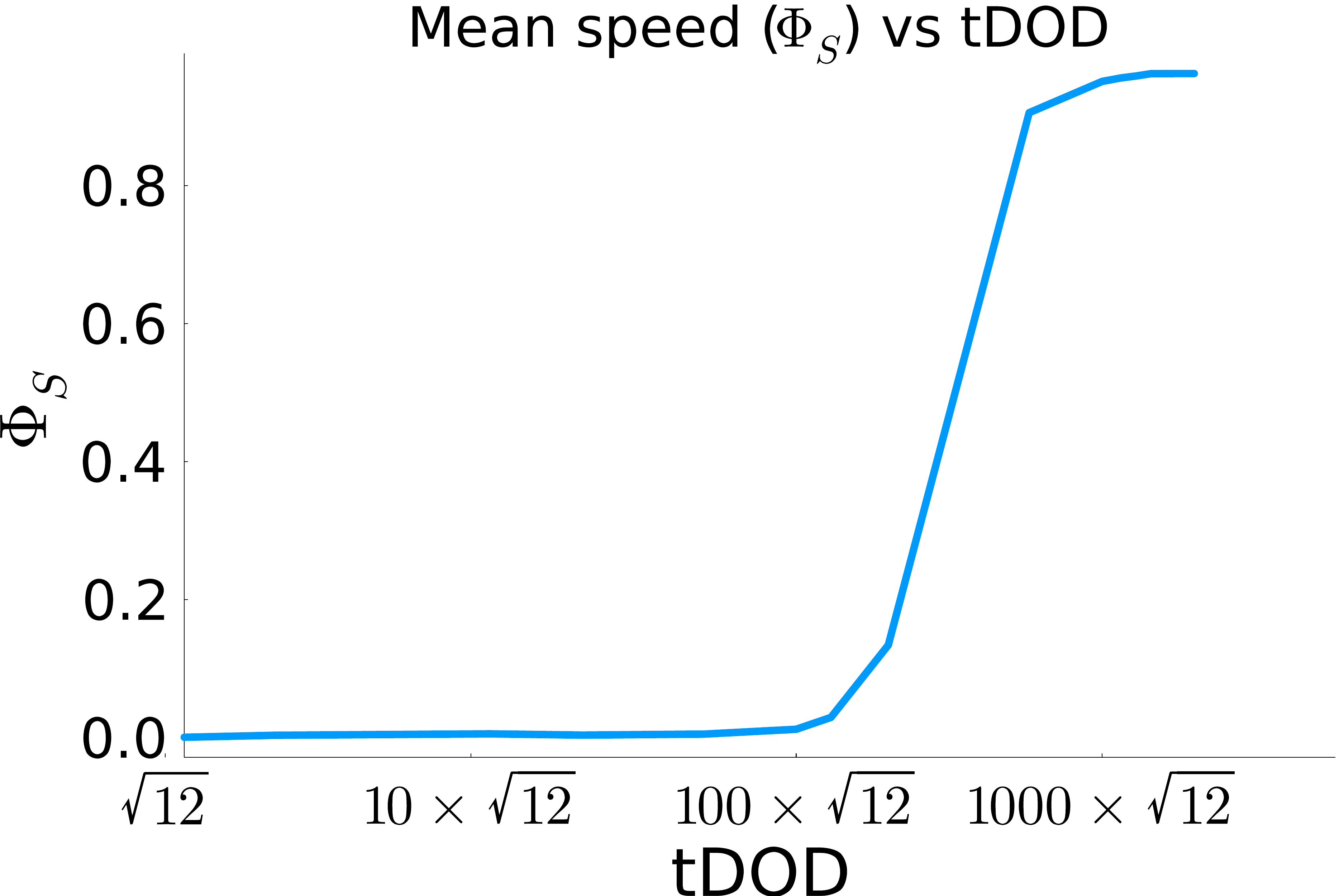}
            \caption{}
            \label{fig:ms_vs_tdod}
        \end{subfigure}
        \caption{Figure (a) and (b) depict the behaviour of the rotational and mean speed order parameters respectively, in response to increases in the target DOD.The phase transition between a crystallised (non-rotating) and rotating flock is clearly visible as the target DOD is increased. Note that a logarithmic scale of base 3 is used.}
        \label{fig:big_ones}
    \end{figure}

Our simulations show that the level of continued movement and collective order in the flock is dependent on the DOD area that the agents target - the target DOD must be greater than some threshold for further collective motion to occur. In fact, as shown in figures \ref{fig:rot_o_alt_vs_tdod} and \ref{fig:ms_vs_tdod}, the response of the flock's collective order with respect to changes to the target DOD bears similarities with phases changes between the states of matter. Illustrations of how the flock behaves in what we call the three main ``phases" can also be seen in supplementary videos 2, 3, and 4. 


From the graphs, we see that for a target DOD on the order low $100$s, the flock remains in what we call the ``\textit{crystallised}" phase, as the agents will continue to form tight, non-moving clusters found in \cite{ALGAR201982} Analysing the lack of movement from an individual agents' perspective, the lack of continued motion occurs due to the fact that any potential positions that are better than an agents’ current position lie within the interior of the flock - however, due to the compactness of the flock, such positions are invalid due to volume exclusion, as illustrated in figure \ref{fig:crystal_state_sample}. Due to the low mean speed of the flock, it naturally follows that the rotational order is also low. 

\begin{figure}[H]
        \centering
        \begin{subfigure}[t]{0.36\textwidth}
         \centering
         \includegraphics[width=\textwidth]{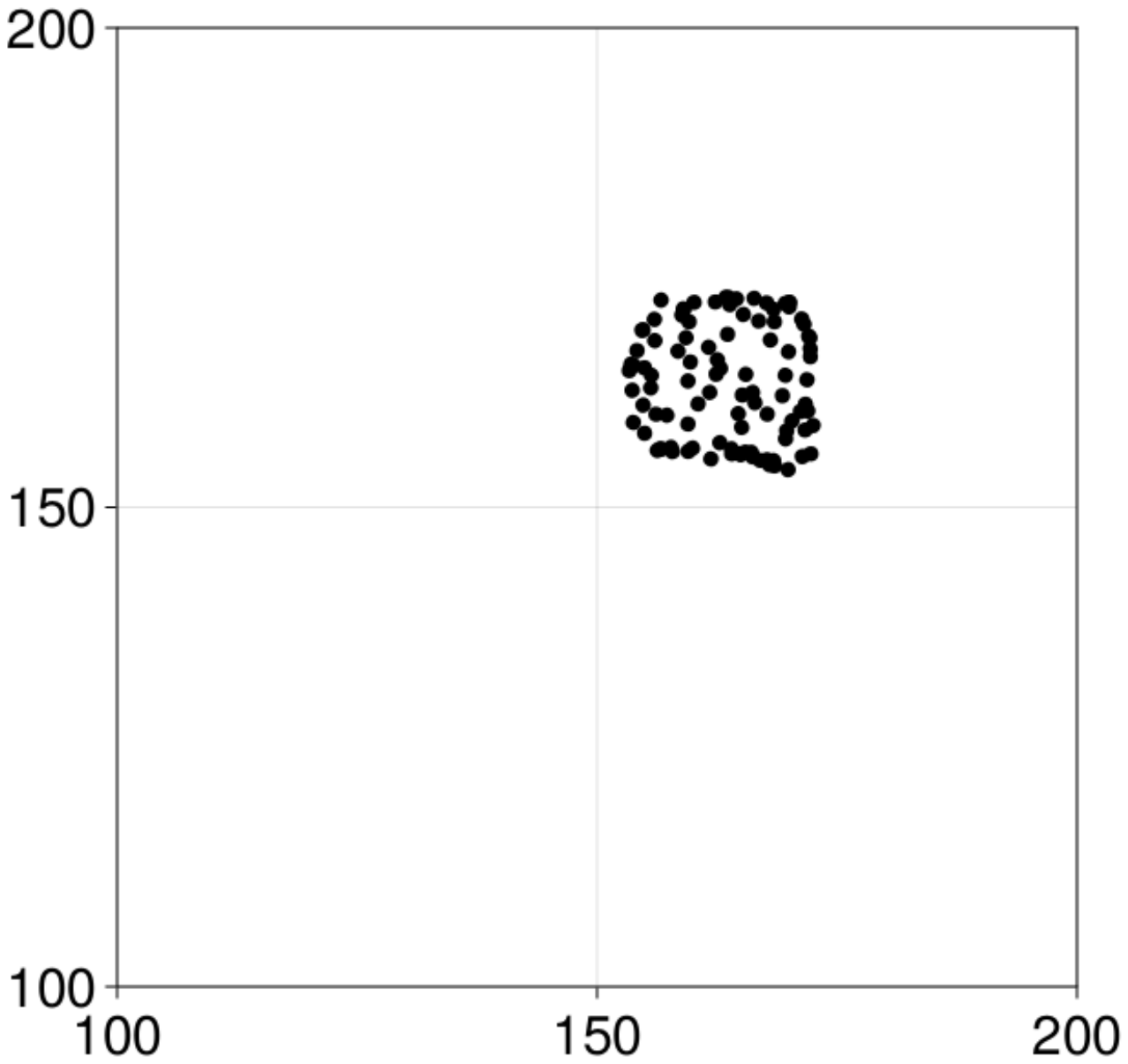}
         \caption{}
         \label{fig:crystal}
     \end{subfigure}
     \hspace{-1cm}
        \begin{subfigure}[t]{0.36\textwidth}
         \centering
         \includegraphics[width=\textwidth]{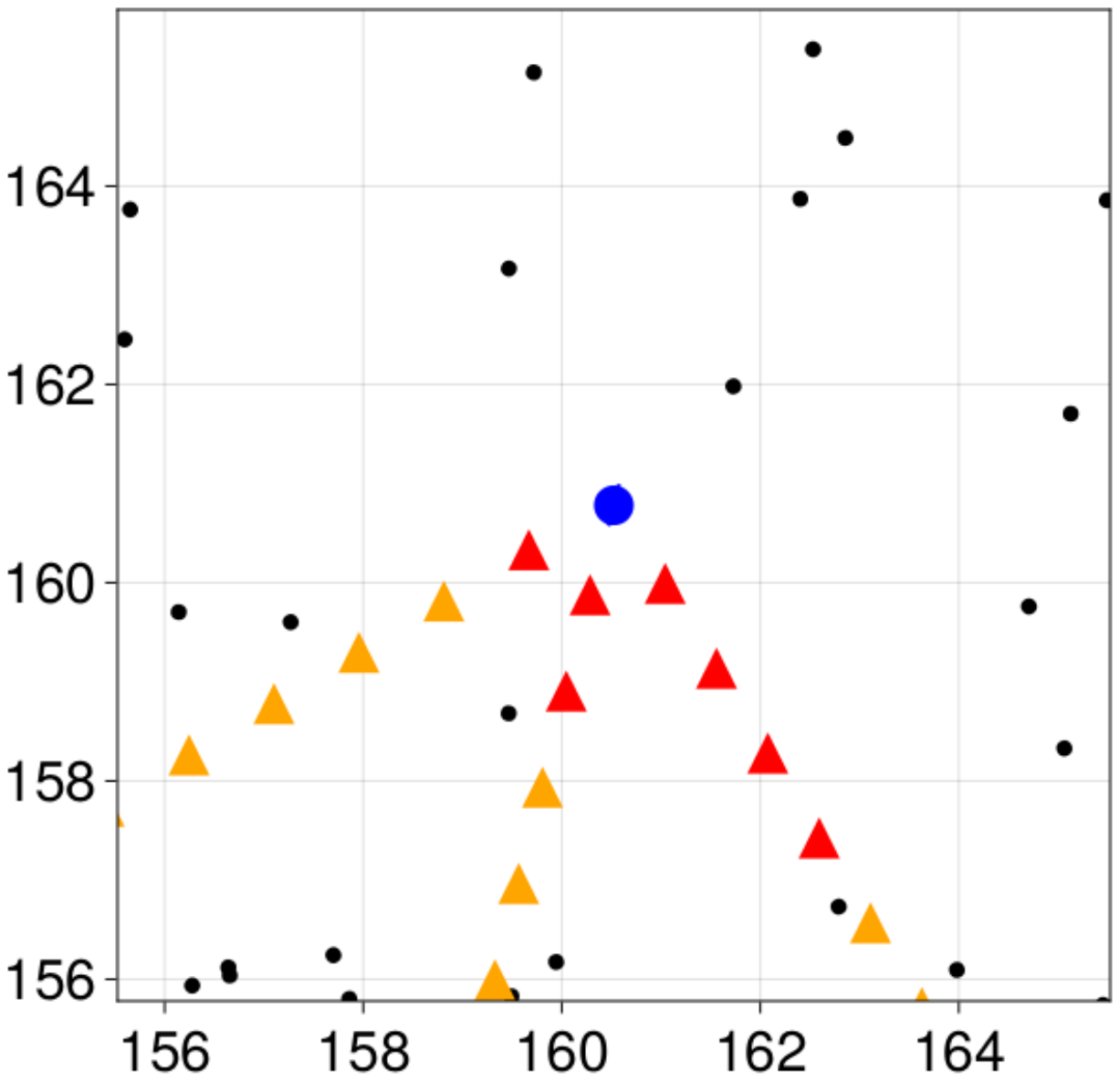}
         \caption{}
         \label{fig:crystal_sample_out}
     \end{subfigure}
     \hspace{-1cm}
     \begin{subfigure}[t]{0.36\textwidth}
         \centering
         \includegraphics[width=\textwidth]{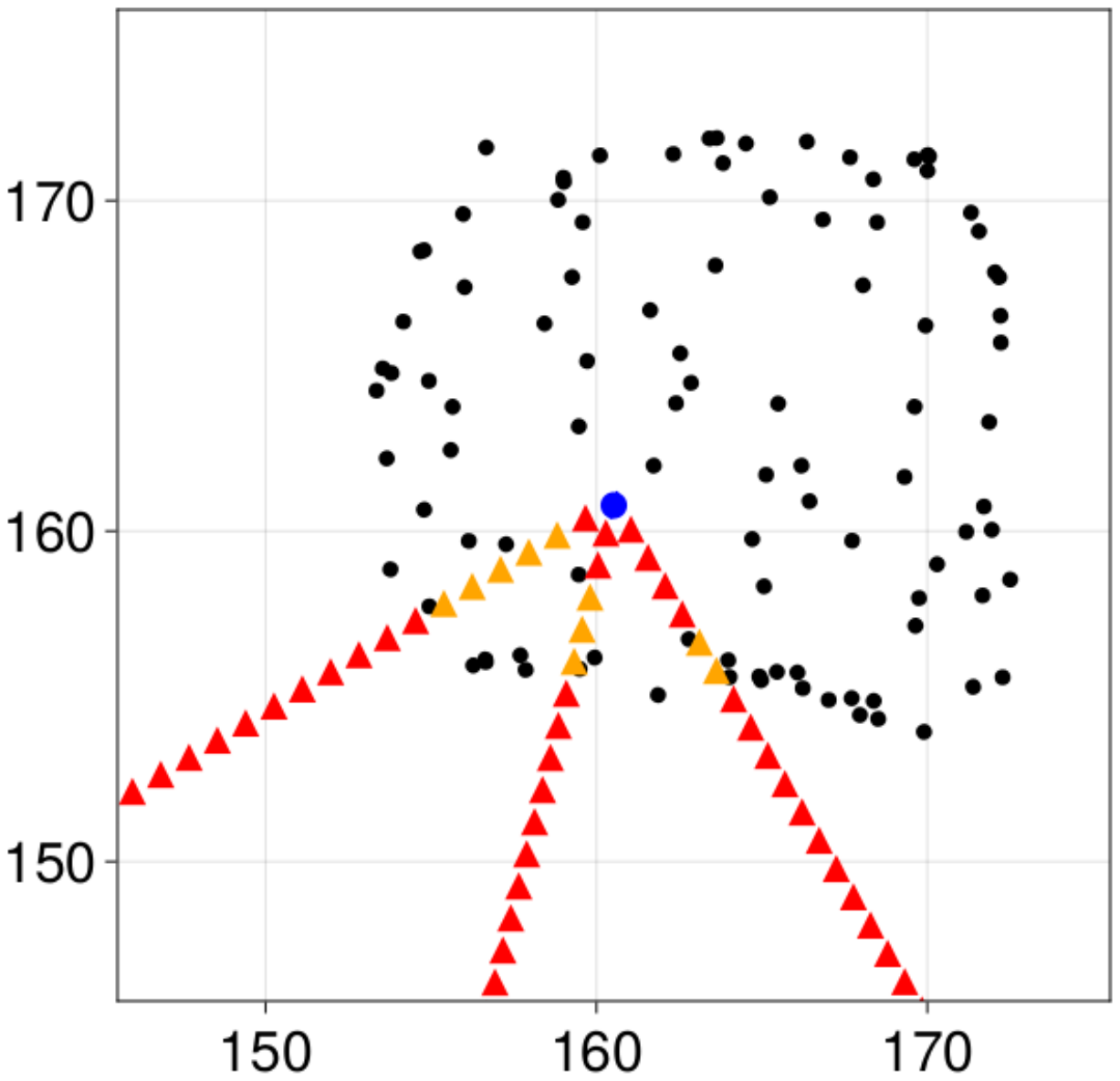}
         \caption{}
         \label{fig:crystal_sample}
     \end{subfigure}
     \caption{(a) An illustration of the flock formed by the agents in a simulation of the crystallised phase. (b) An illustration of some of the positions sampled by an agent in the flock. The agent's current position is indicated by the blue dot, while the sampled positions are triangles. The colour of the triangle indicates the suitability of the position - red if the position results in a domain further from the tDOD than the agent's current position, orange if the potential area is closer to the tDOD but is invalid due to volume exclusion, and green if the potential area is closer to the target area and the position is valid. We see that while there are no green triangles, there are quite a few orange ones. This indicates that while there are better positions within the flock for the agent to travel to, these positions are all excluded due to volume exclusion. (c) Illustrates the sampling of the same agent from further out - we see that the agent has no better positions to move to because the only positions that have better areas are inside the flock but are invalid due to volume exclusion, while areas  outside the flock are poor as they have too large an area.}
    \label{fig:sample_ill_liquid}
    \end{figure}

\begin{figure}[H]
        \centering
        \begin{subfigure}[t]{0.36\textwidth}
         \centering
         \includegraphics[width=\textwidth]{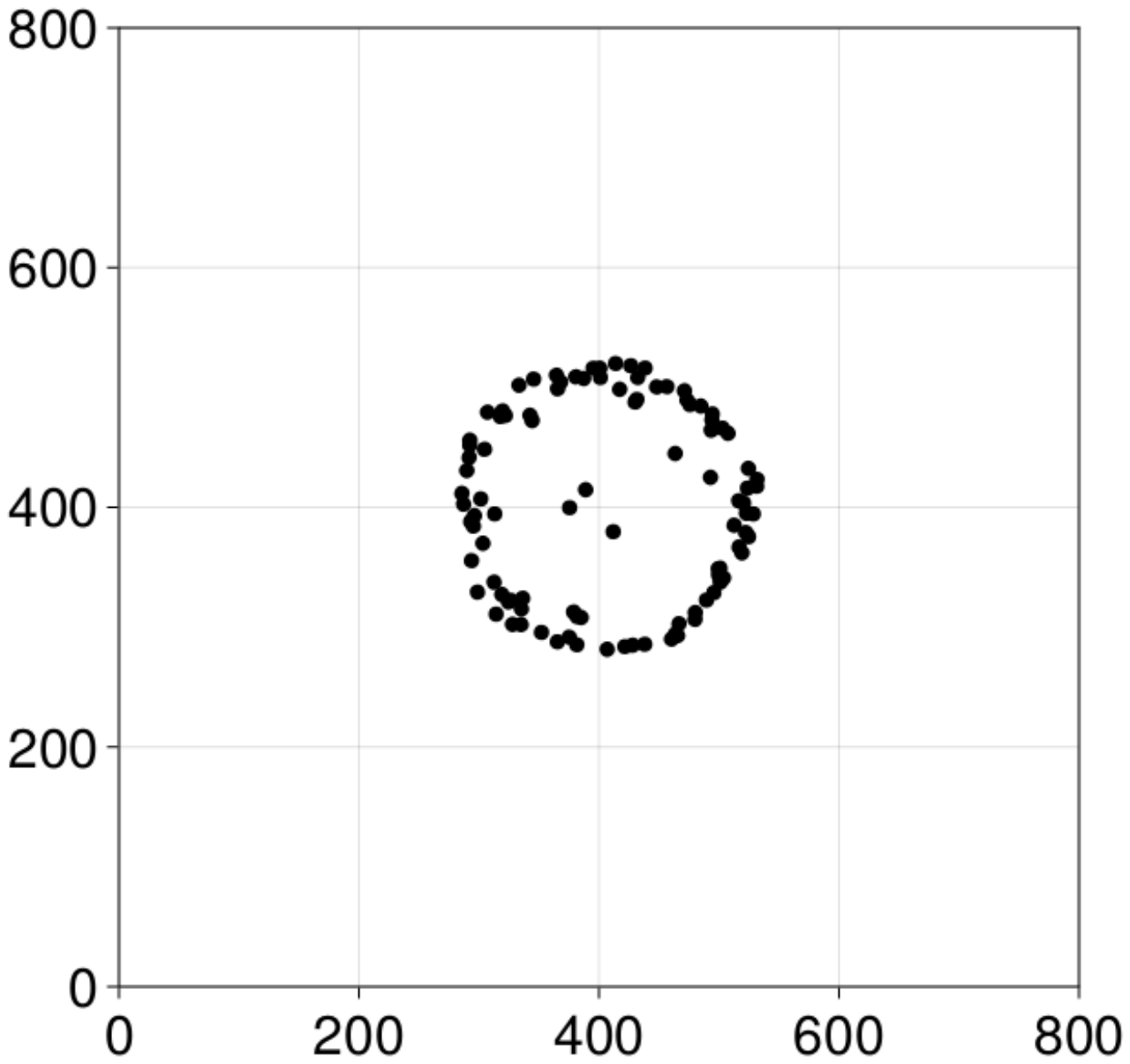}
         \caption{}
         \label{fig:circle_expansion}
     \end{subfigure}
    \hspace{-1cm}
     \begin{subfigure}[t]{0.36\textwidth}
         \centering
         \includegraphics[width=\textwidth]{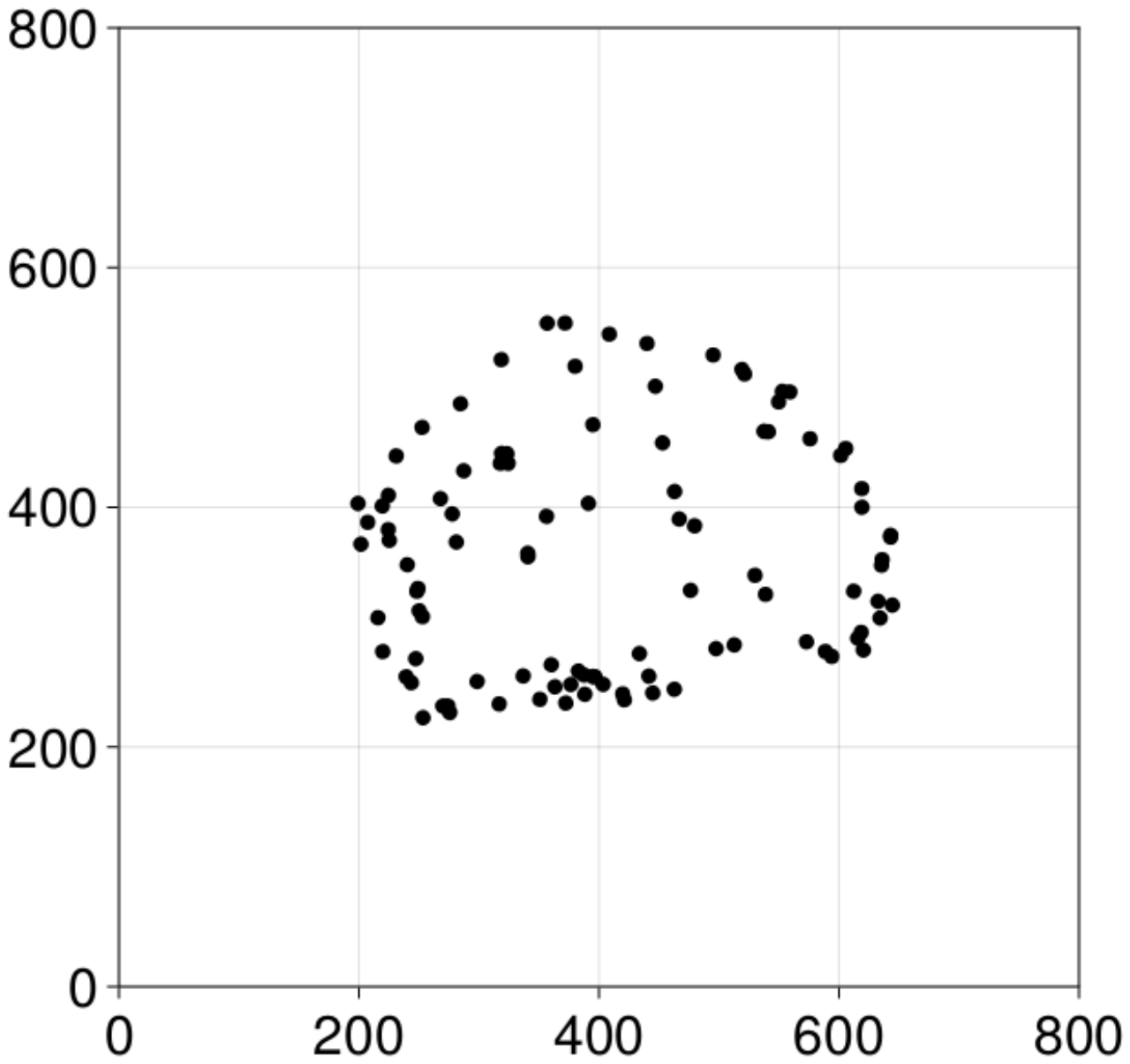}
         \caption{}
         \label{fig:i19_no_title_close}
     \end{subfigure}
    \hspace{-1cm}
     \begin{subfigure}[t]{0.36\textwidth}
         \centering
         \includegraphics[width=\textwidth]{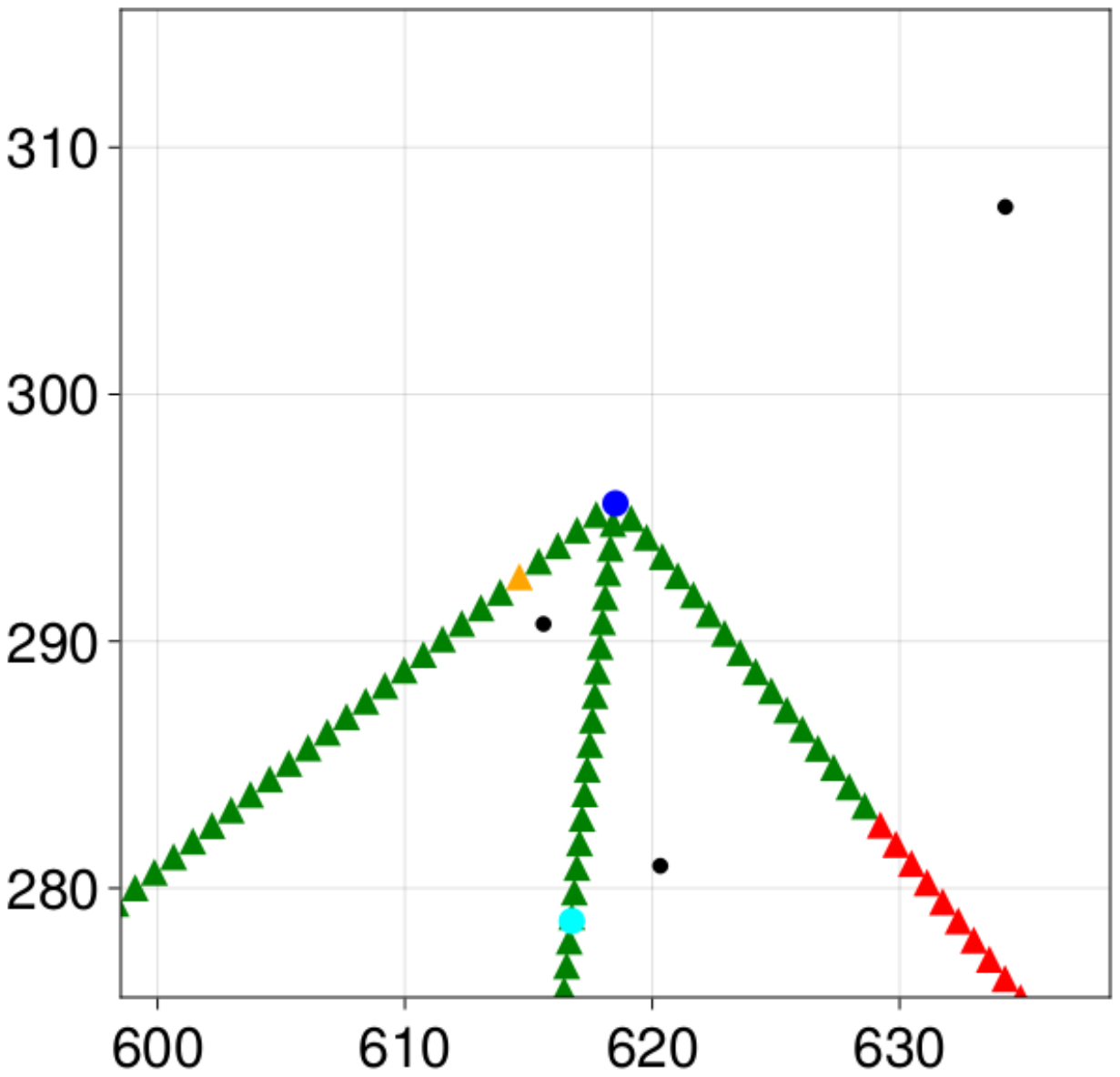}
         \caption{}
         \label{fig:stdod25_show_move_closest}
     \end{subfigure}
     \caption{(a) An illustration of the circular formation adopted by the flock in the initial expansion period for larger tDODs - in this case, a tDOD of $1000\times\sqrt{12}$. (b) An example of the non-stationary equilibrium state the flock reaches after the outward circular expansion. (c) illustrates the positions sampled by an agent in the flock after equilibrium is established using the same colouring scheme as in the previous figure. We can see that there are many more optimal positions (green positions) in comparison to the sampled positions in a crystallised state, with the optimal position shown in cyan.}
    \label{fig:sample_ill_liquid}
    \end{figure}

As the tDOD rises above a value in the range of $100\times\sqrt{12}$ to $2000\times\sqrt{12}$ however, the flock enters what we call the ``\textit{liquid}" phase, and the mean speed of the agents begins to increase. From the perspective of the individual agents, the increasing mean speed of the flocks occurs due to the fact that the increasing flock size leaves inter-agent distances greater. This decreases the number of optimal points that would otherwise have been excluded due to volume exclusion rules in the solid flock phase as illustrated in Figure \ref{fig:i19_no_title_close}. A greater number of agents are therefore able to move to more optimal positions at every time step. For this range of tDODs, one common behaviour is that the flock will in fact expand outwards from its initial configuration in a circular formation with the majority of agents located on (or close to) the perimeter of the flock, while the interior is sparsely populated. Eventually however, more agents begin to fill in the interior of the flock, and the circular formation deforms into more irregular and unordered formation.

\begin{figure}[ht]
        \begin{subfigure}[t]{0.36\textwidth}
            \centering
            \includegraphics[width=\textwidth]{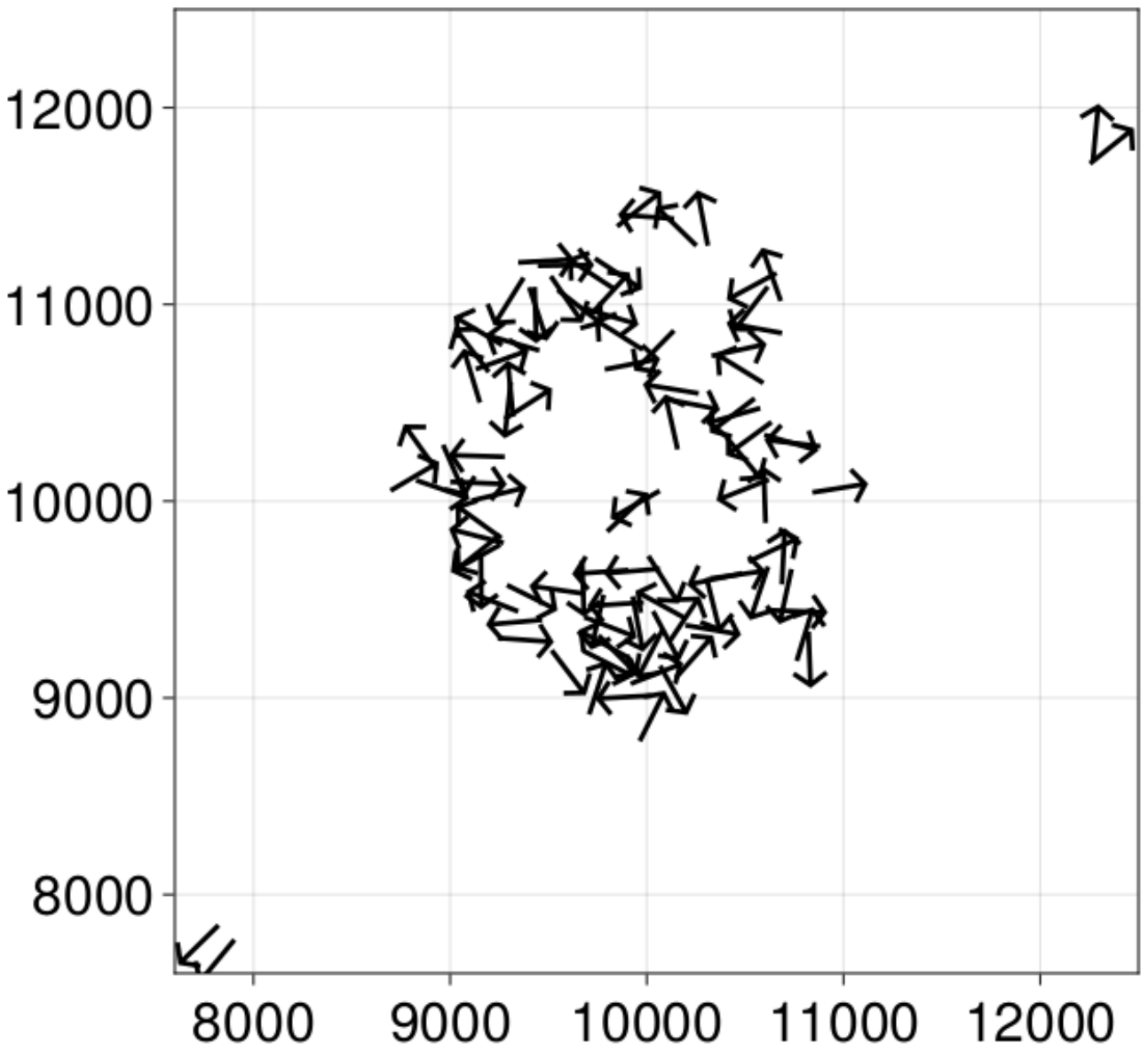}
            \caption{}
        \end{subfigure}
        \hspace{-1cm}
        \begin{subfigure}[t]{0.36\textwidth}
            \includegraphics[width=\textwidth]{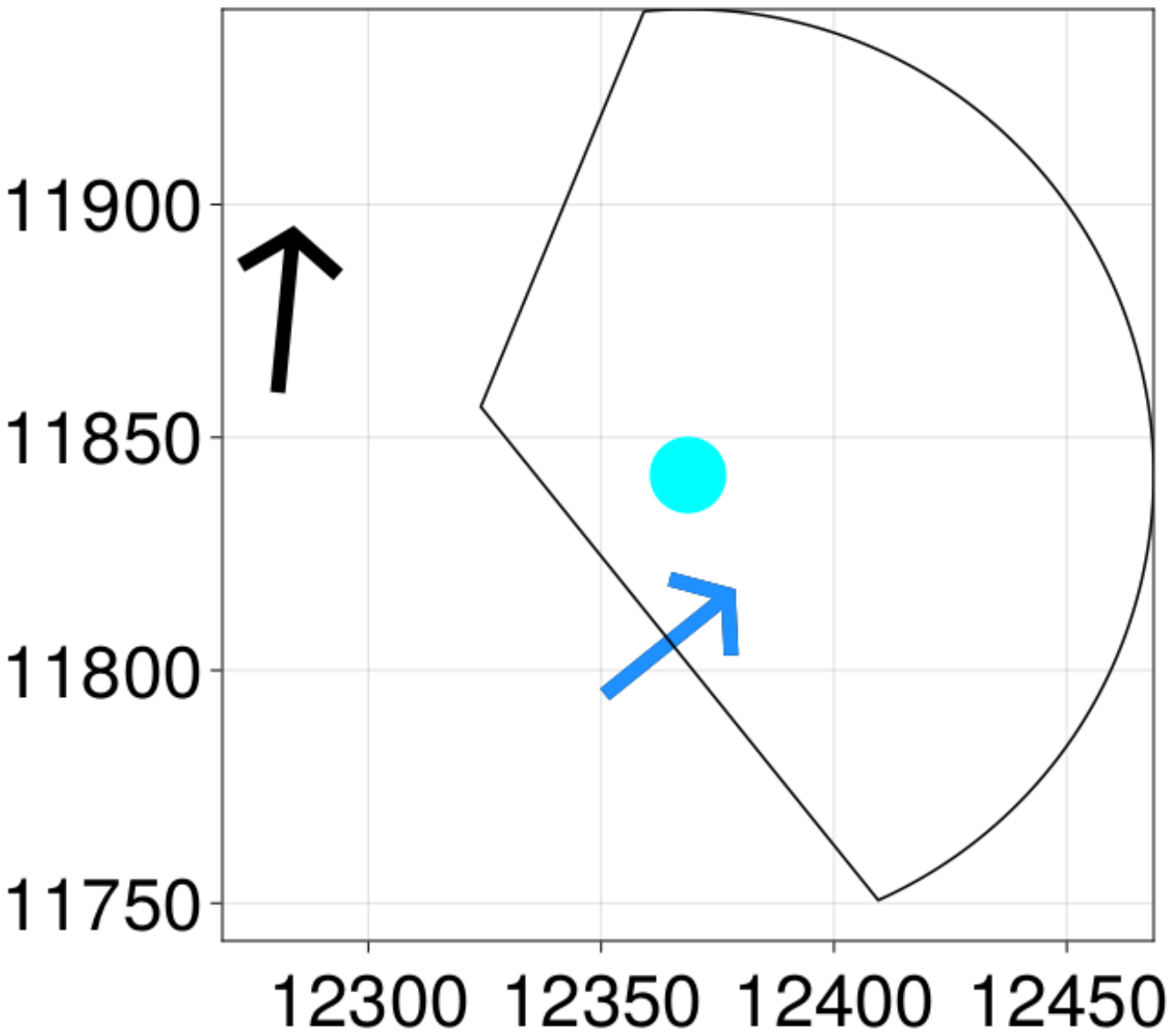}
            \caption{}
        \end{subfigure}
        \hspace{-1cm}
        \begin{subfigure}[t]{0.36\textwidth}
            \includegraphics[width=\textwidth]{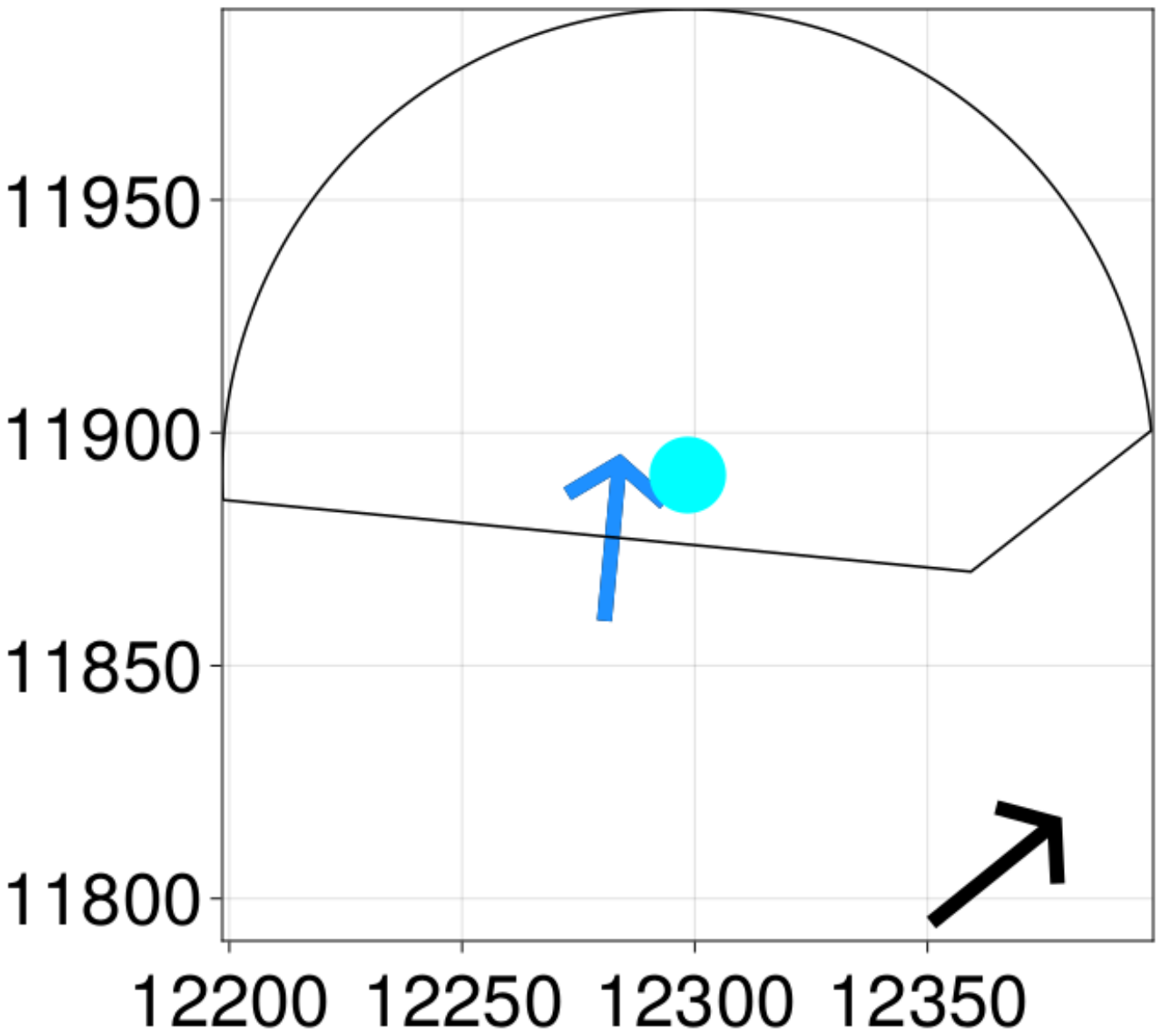}
            \caption{}
        \end{subfigure}
        \caption{(a) An illustration of the pairs flying behaviour in a flock where agents have a target DOD of 18000. Note that the domain size is $20000\times20000$. The full video of this simulation is given by supplementary video 5. (b) and (c) give a closer view of the agents flying away in the top right hand area of figure a. The cyan dot in each figure indicates the ideal position for the agent in light blue. The figures illustrate how the pairs of agents leaving the main flock are able to maintain motion -- combined with the fact that their DOD is forward-bounded, each agent is able to use the half plane from their partner to bound an otherwise unbounded DOD. } 
        \label{fig:sample_evap}
    \end{figure}

Once the flock tDOD exceeds $2000\times\sqrt{12}$ however, while the mean speed and initial expansion of the flock are similar to that of the liquid phase, the eventual non-stationary equilibrium state of the flock is much less ellipsoid in shape. 
Moreover, with an increasing tDOD, agents are more likely to choose a purely radial movement relative to the flock’s center of mass as they seek to increase their DOD by increasing the distance to their neighbours. Thus, above a tDOD of $2000\times\sqrt{12}$, the rotational order of the flock peaks and actually decreases, resulting in the decrease in the rotational order as seen in figure \ref{fig:rot_o_alt_vs_tdod}. 
One particularly interesting behaviour that became increasingly prominent with higher tDODs was what we have termed as ``pair flying": where agents would break away from the main flock altogether to travel in pairs, as seen in figure \ref{fig:sample_evap}. 

\begin{figure}[H]
        \begin{subfigure}[t]{0.5\textwidth}
            \includegraphics[width=\textwidth]{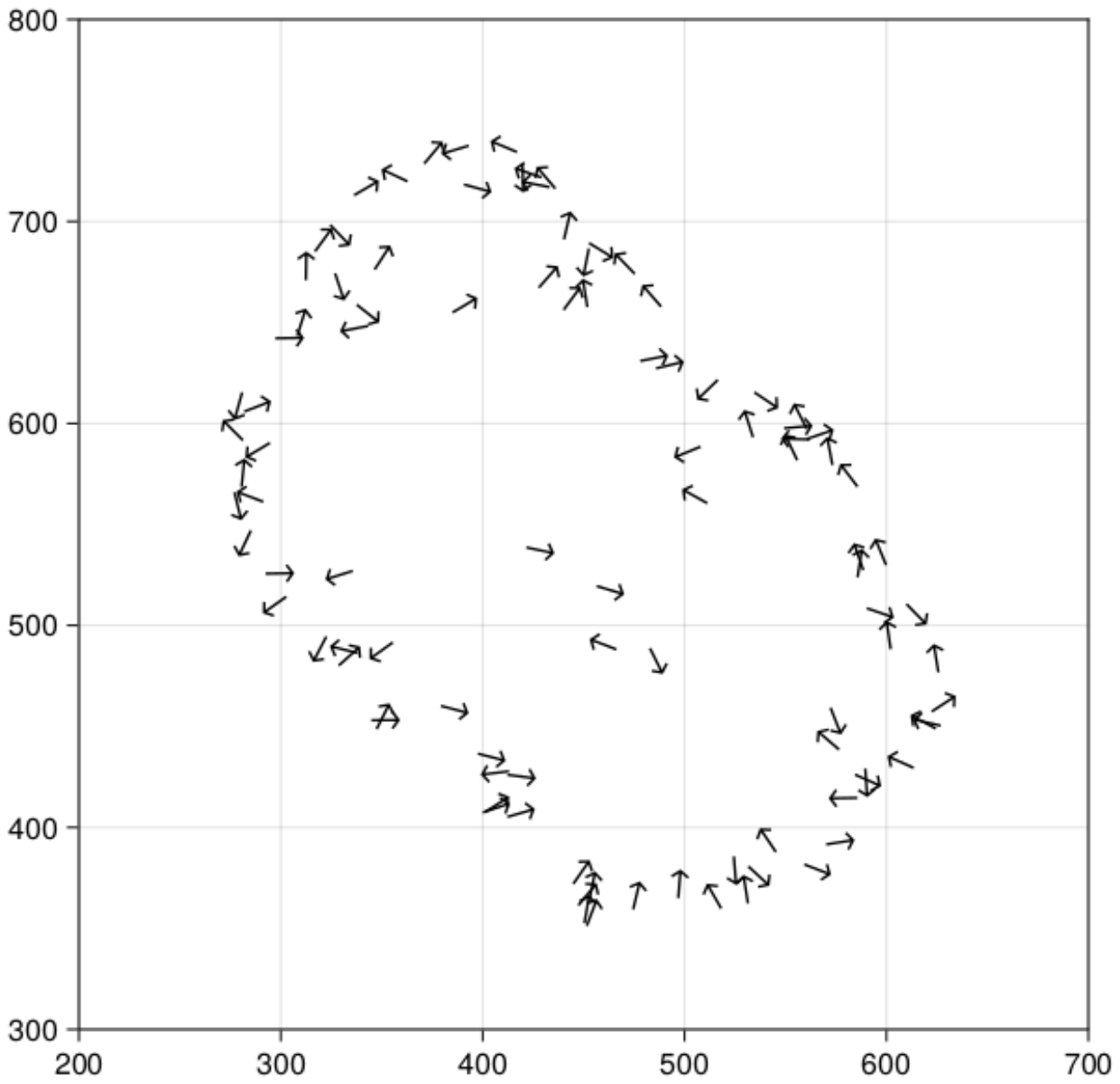}
            \caption{}
            \label{fig:cave_in_space}
        \end{subfigure}
        \begin{subfigure}[t]{0.5\textwidth}
            \includegraphics[width=\textwidth]{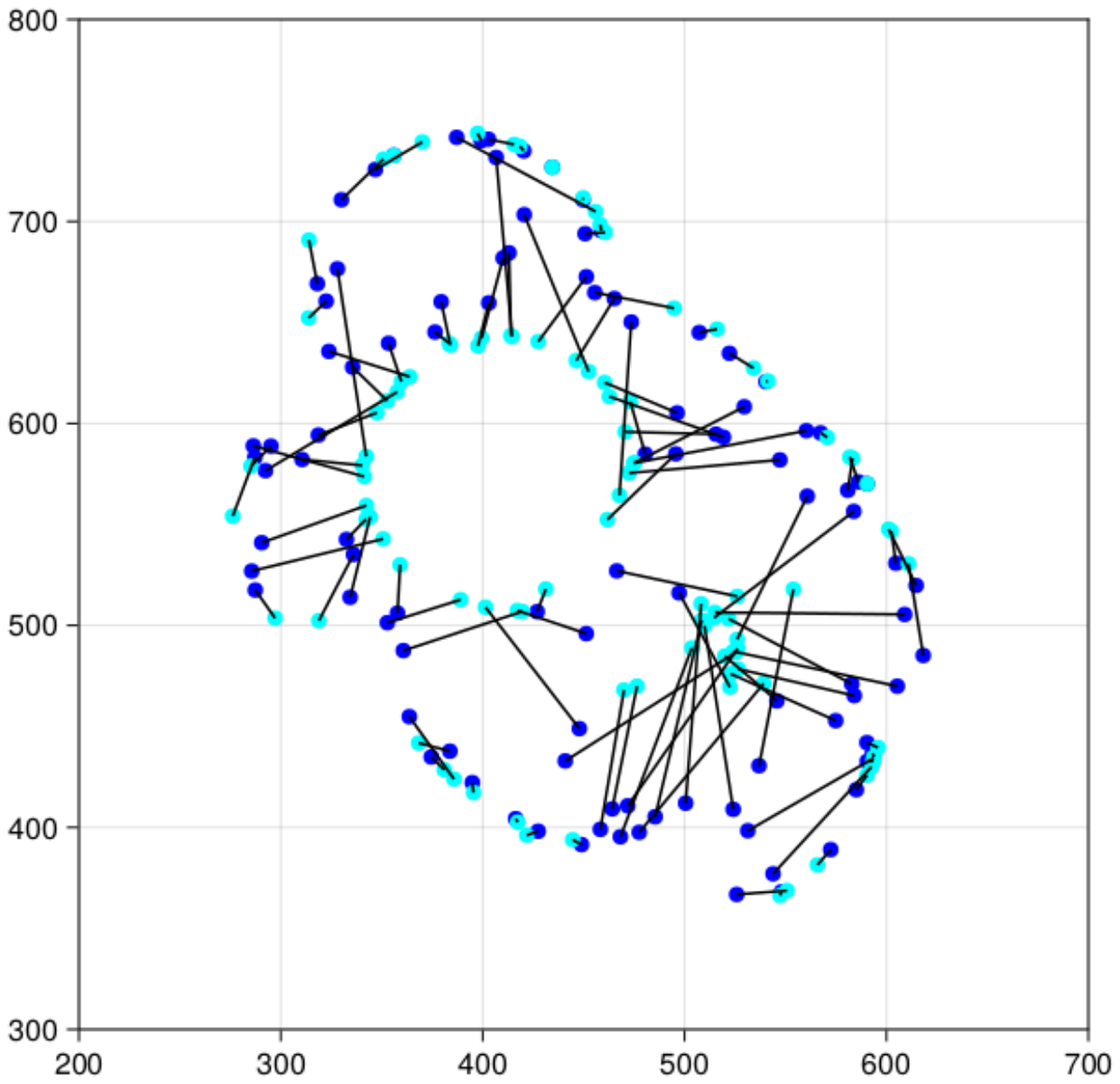}
            \caption{}
            \label{fig:cave_in_DOD}
        \end{subfigure}
        \caption{An illustration of the cave-in effect, from the simulation in supplementary video 6. Preceding the cave-in behaviour, a large empty space opens on the interior of the flock, such as the one in figure \ref{fig:cave_in_space} around the point of $(350.0, 600.0)$. Figure \ref{fig:cave_in_DOD} illustrates the ideal positions agents aim for. Agent positions are shown as the dark blue dots, while the ideal position for each agent is shown as the cyan dots linked to the original positions by a black line. We see that many agents have their ideal position on the edge of the ``void" in the middle of the flock.}
        \label{fig:cave_in_start}
    \end{figure}

\begin{figure}[H]
        \begin{subfigure}[t]{0.5\textwidth}
            \includegraphics[width=\textwidth]{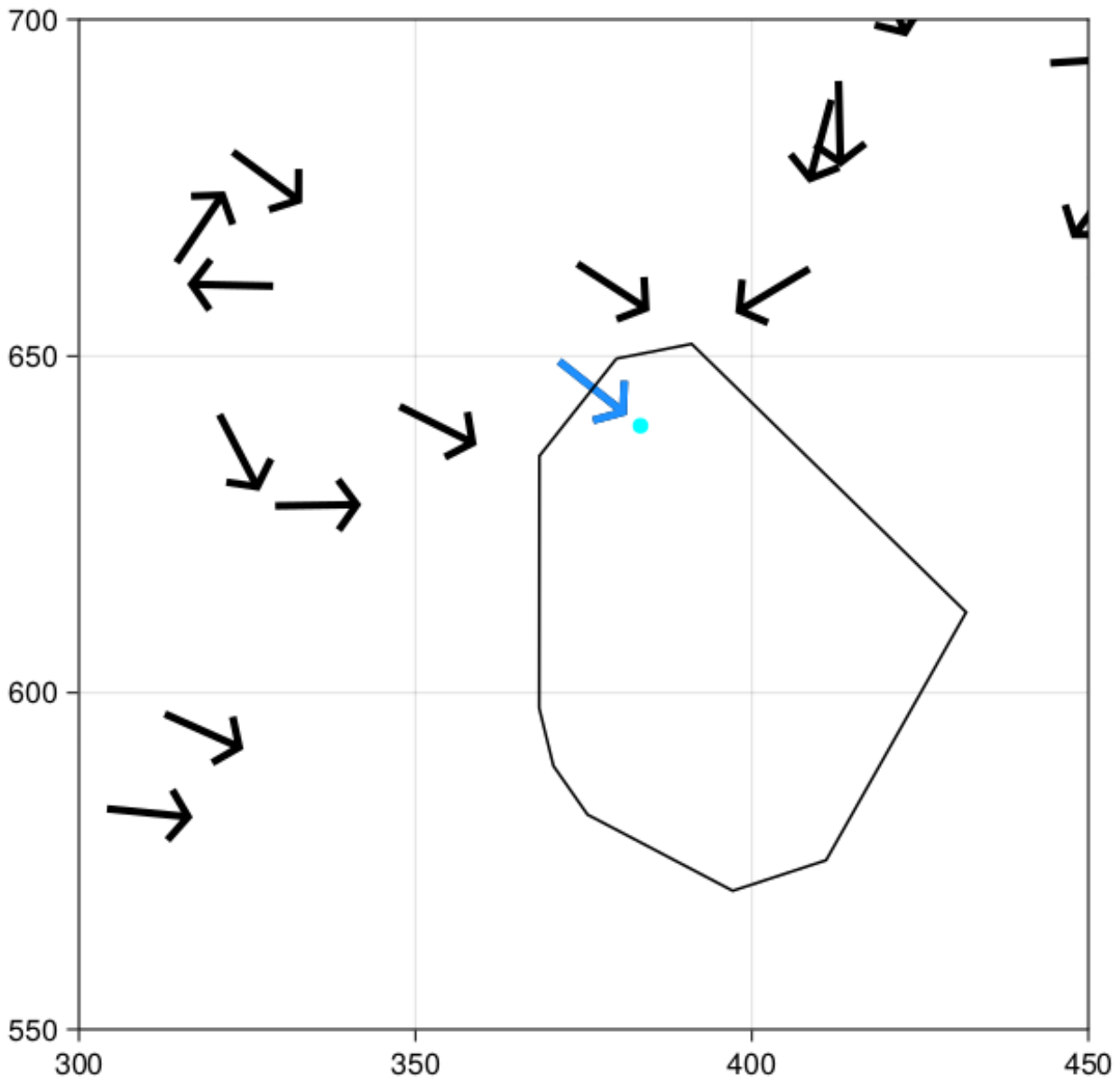}
            \caption{}
            \label{fig:cave_in_desirables}
        \end{subfigure}
        \begin{subfigure}[t]{0.5\textwidth}
            \includegraphics[width=\textwidth]{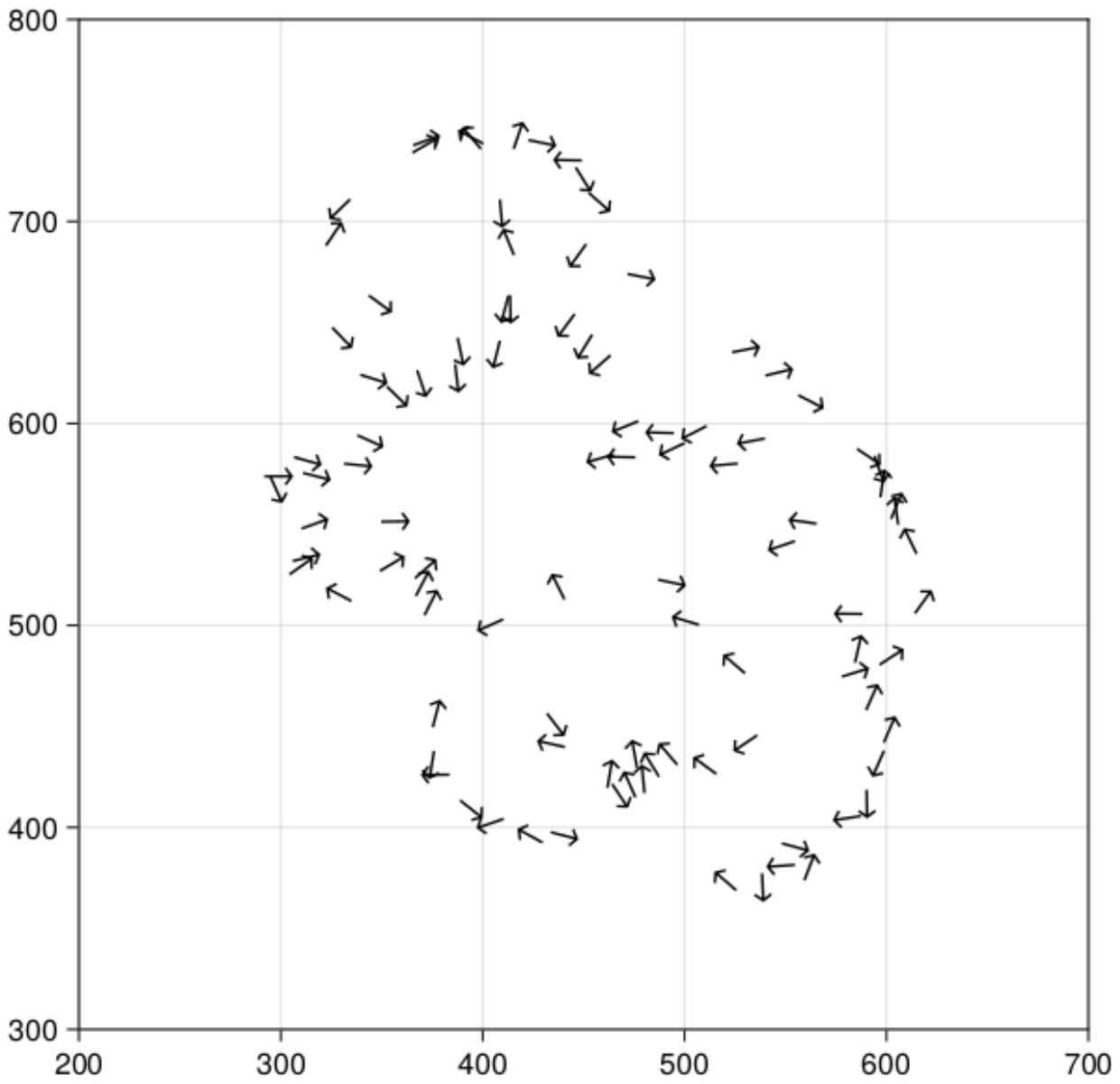}
            \caption{}
            \label{fig:cave_in_convergence}
        \end{subfigure}
        \caption{Figure \ref{fig:cave_in_desirables} Shows the desired position (cyan dot) and potential domain in that position for one of the agents heading towards the ``void", with its current position and direction indicated by the light blue arrow. We see that the domain comprises a large part of the empty space in the flock, but is still bounded by other agents around the agent. Figure \ref{fig:cave_in_convergence} shows the convergence  of the agents on the ``void" of empty space.}
        \label{fig:cave_in}
    \end{figure}

Another behaviour that is observed in this tDOD range is what we call the ``caving-in" effect, where large numbers of agents along the perimeter of the flock converge to a point on the interior of the flock. As shown in figures \ref{fig:cave_in_start} and \ref{fig:cave_in}, this behaviour occurs when a ``void" of empty space forms in the interior of the flock. This space then offers agents highly desirable positions along the edge of the void. The DODs in these positions - illustrated by figure \ref{fig:cave_in_DOD} - are neither too large nor too small, as they occupy a large amount of space from the void, while still being bounded by neighbouring agents on the edge of the void. Thus, the DODs in these positions are very close to the target DOD. 

 Then, since these positions lead to almost ideal DODs, agents which can sample these positions converge on the void. As they do so however, they fill in the spaces that would otherwise have been ideal, and so decrease the number of ideal positions. This occurs until eventually, the originally empty space in the interior is filled, and the agents are forced to find optimal positions elsewhere.

\begin{figure}[H]
        \begin{subfigure}[t]{0.45\textwidth}
            \centering
            \includegraphics[width=\textwidth]{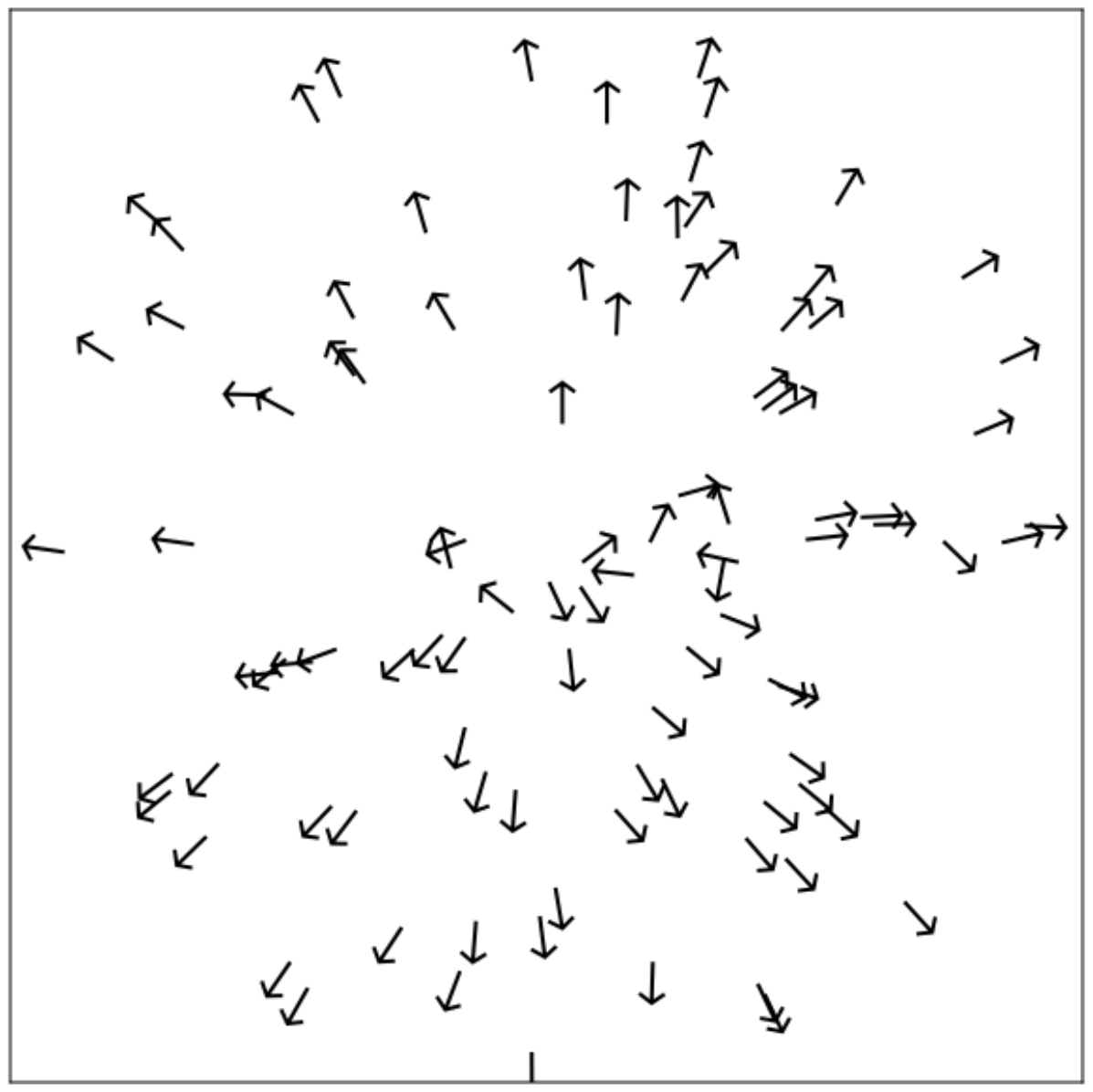}
            \caption{}
        \end{subfigure}
        \begin{subfigure}[t]{0.45\textwidth}
            \includegraphics[width=\textwidth]{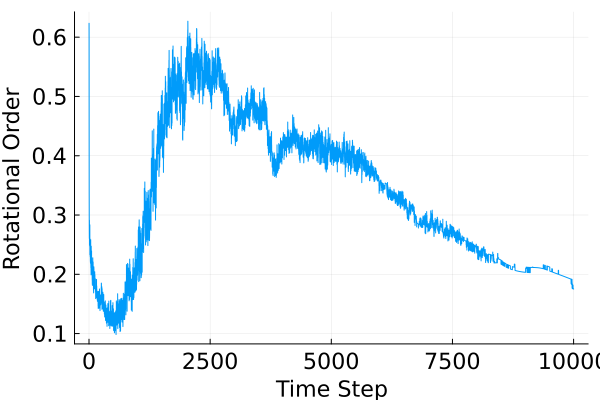}
            \caption{}
        \end{subfigure}
        \caption{(a) An illustration of the flock formation at time step 10000 when agents have a tDOD $\ge 22000$. Shown is tDOD$=2200$  leading to almost purely radial movement. Note that the domain size of the simulation in this case is $20000\times20000$. (b) An illustration of the low rotational order obtained with large tDODs - indicating that most agent movement is dedicated to radial movement instead. } 
        \label{fig:less_ellipsoid}
    \end{figure}

Eventually, as the tDOD reaches a level above 22000, the flock enters what we call the ``\textit{gaseous}" phase, where the movement of the agents becomes purely radial, as shown in figure \ref{fig:less_ellipsoid} and supplementary video 4. 

\subsection{Discussion}
\begin{figure}[H]
      \centering
        \begin{subfigure}[t]{0.32\textwidth}
         \centering
         \includegraphics[width=\textwidth]{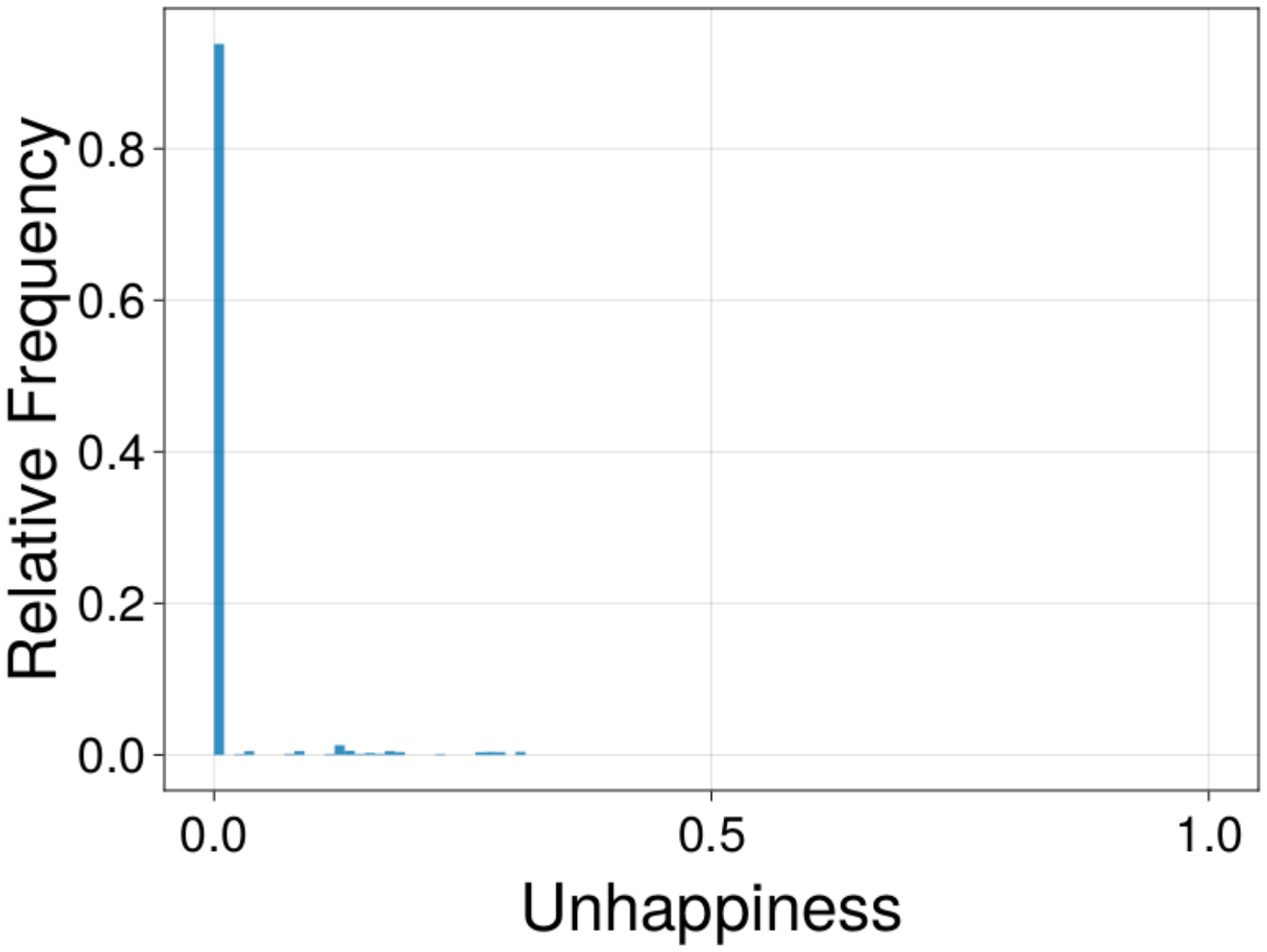}
         \caption{Crystallised phase unhappiness distribution}
         \label{fig:stdod41_hist}
     \end{subfigure}
     \begin{subfigure}[t]{0.32\textwidth}
         \centering
         \includegraphics[width=\textwidth]{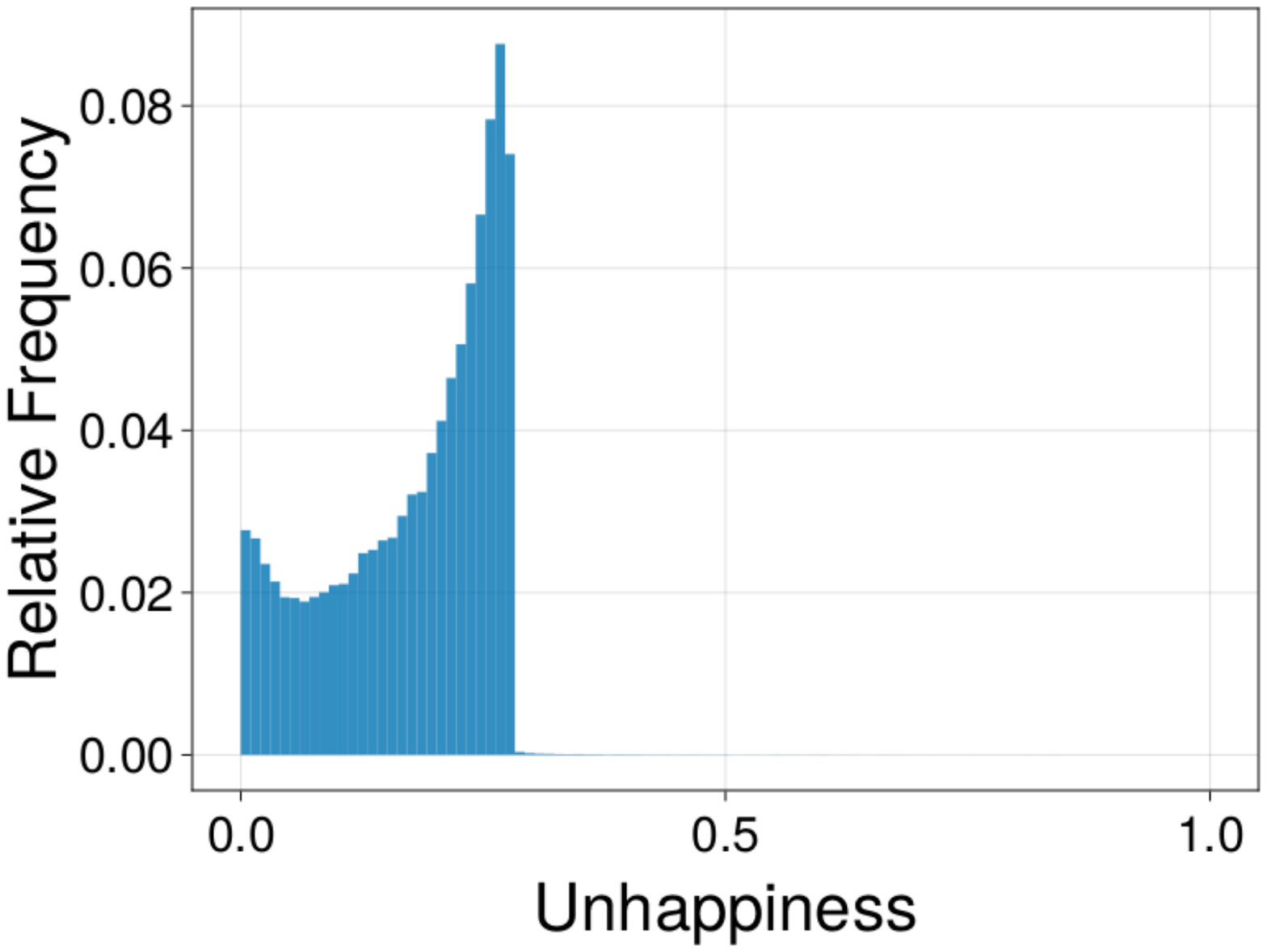}
         \caption{Liquid phase unhappiness distribution}
         \label{fig:stdod42_hist}
     \end{subfigure}
     \begin{subfigure}[t]{0.32\textwidth}
         \centering
         \includegraphics[width=\textwidth]{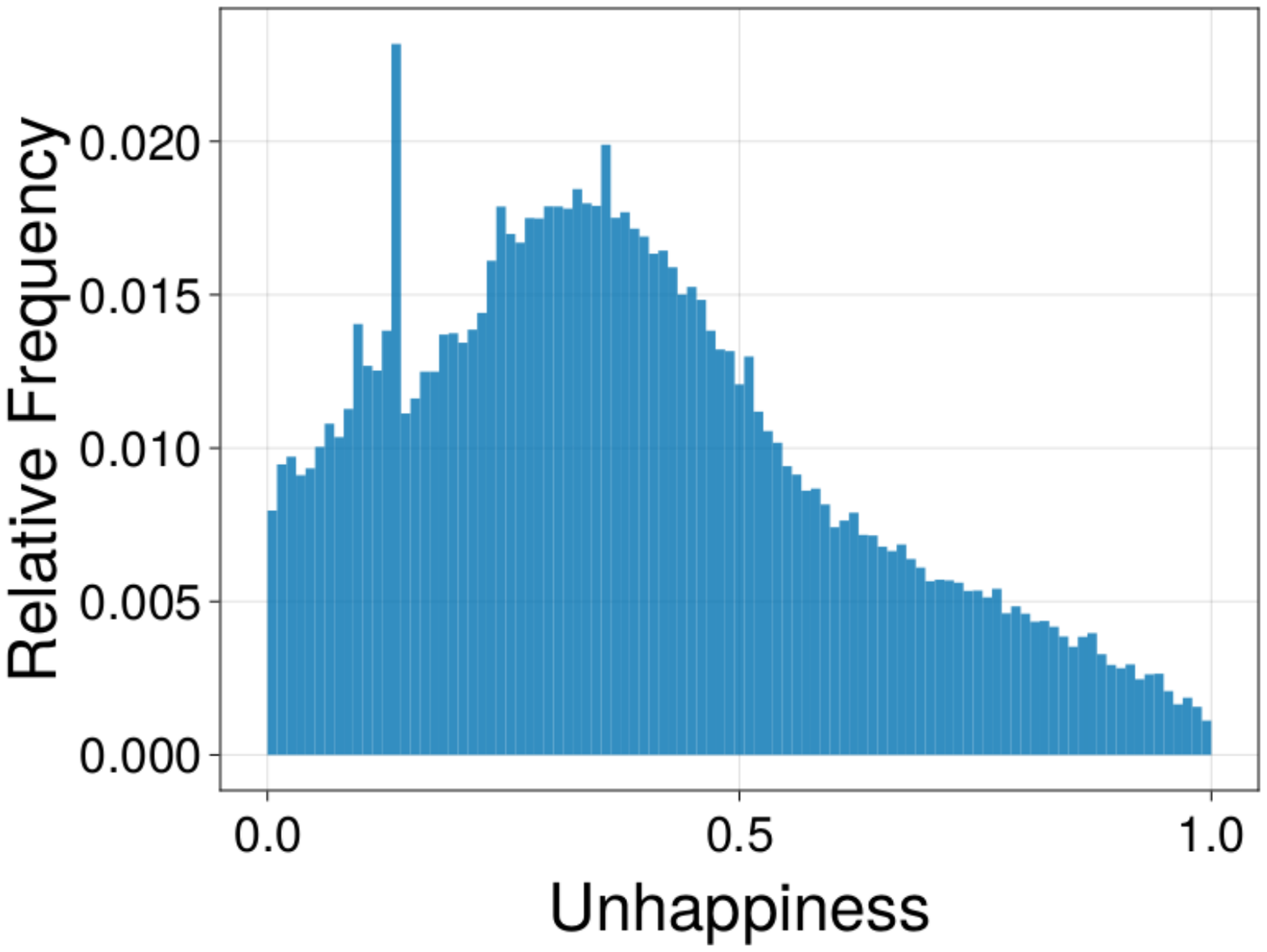}
         \caption{Gaseous phase unhappiness distribution.}
         \label{fig:stdod43_hist}
     \end{subfigure}
     \caption{Probability distribution of the agent unhappiness. (a) The unhappiness distribution for a flock with tDOD of 0.0. (b) The unhappiness distribution for a flock with target DOD of $1000\times\sqrt{12}$. (c) The unhappiness distribution for a flock with target DOD of $4000\times\sqrt{12}$.}
\end{figure}

One way we can confirm the phase changes in a quantitative manner is to measure the level of order/disorder in the system. 
To do so, we draw upon a measure of order from statistical mechanics: entropy. In statistical mechanics, the entropy of a system is proportional to the number of states the system can be in. If we were to think about the system's state probability distribution, the entropy is proportional to the spread of that distribution. The number of states for a system of identical particles is proportional to the number of possible single particle states that can be occupied. The state of a single particle in turn can be specified by the value it has for some physical observable such as energy. Analogously, we posit that we can measure the entropy of our system by looking at the distribution of a quantity we call the ``\textit{unhappiness}” for each agent. The unhappiness of each agent measures the deviation of the agents’ current DOD against their target DOD, given by

    \begin{align}
        \Delta = \frac{\mathrm{abs}(A_i - A_\mathrm{t})}{\Delta_{\mathrm{max}}},
    \end{align}
where $A_i$ represents the bounded DOD of the $i$-th agent, $A_\mathrm{t}$ represents the target DOD, and $\Delta_{\mathrm{max}}$ represents the largest possible deviation from the target DOD, given by
    \begin{align}
        \Delta_{\mathrm{max}} := \mathrm{max}(|A_\mathrm{t}|, |A_t-0.5\pi\rho^2|).
    \end{align}
Thus, to characterise the phases in our model, we ran a simulation of 5000 steps in each of the three phases, and recorded the unhappiness distribution for each. Specifically, we let the model reach a steady state after 2000 time steps, then recorded the unhappiness levels of each agent for the remaining 3000 time steps. Measuring the possible states of unhappiness the agents can experience in each simulation, we see that each phase discussed previously is characterised by a distinct distribution in the unhappiness of the agents. From figure \ref{fig:stdod41_hist}, we can see that the possible states of unhappiness for individual agents are almost exclusively concentrated in a single bracket of unhappiness. As we increase the tDOD in the liquid phase from $100\times\sqrt{12}$ to $2000\times\sqrt{12}$ however, we find that the spread of unhappiness states increases drastically, before going close to 0 abruptly at some maximum value, as shown in figure \ref{fig:stdod42_hist}. The spread and unhappiness bracket at which probabilities abruptly fall off steadily increases with the increasing tDOD. However, as illustrated in figure \ref{fig:stdod43_hist}, once we reach the gaseous phase, the distribution spread increases significantly once again, and without the abrupt fall-off.

 Importantly, we also note that in the liquid and gaseous phases, the number of agents occupying a given unhappiness level also varies through each time step. 

    \begin{figure}[H]
        \centering
        \includegraphics[width=0.5\textwidth]{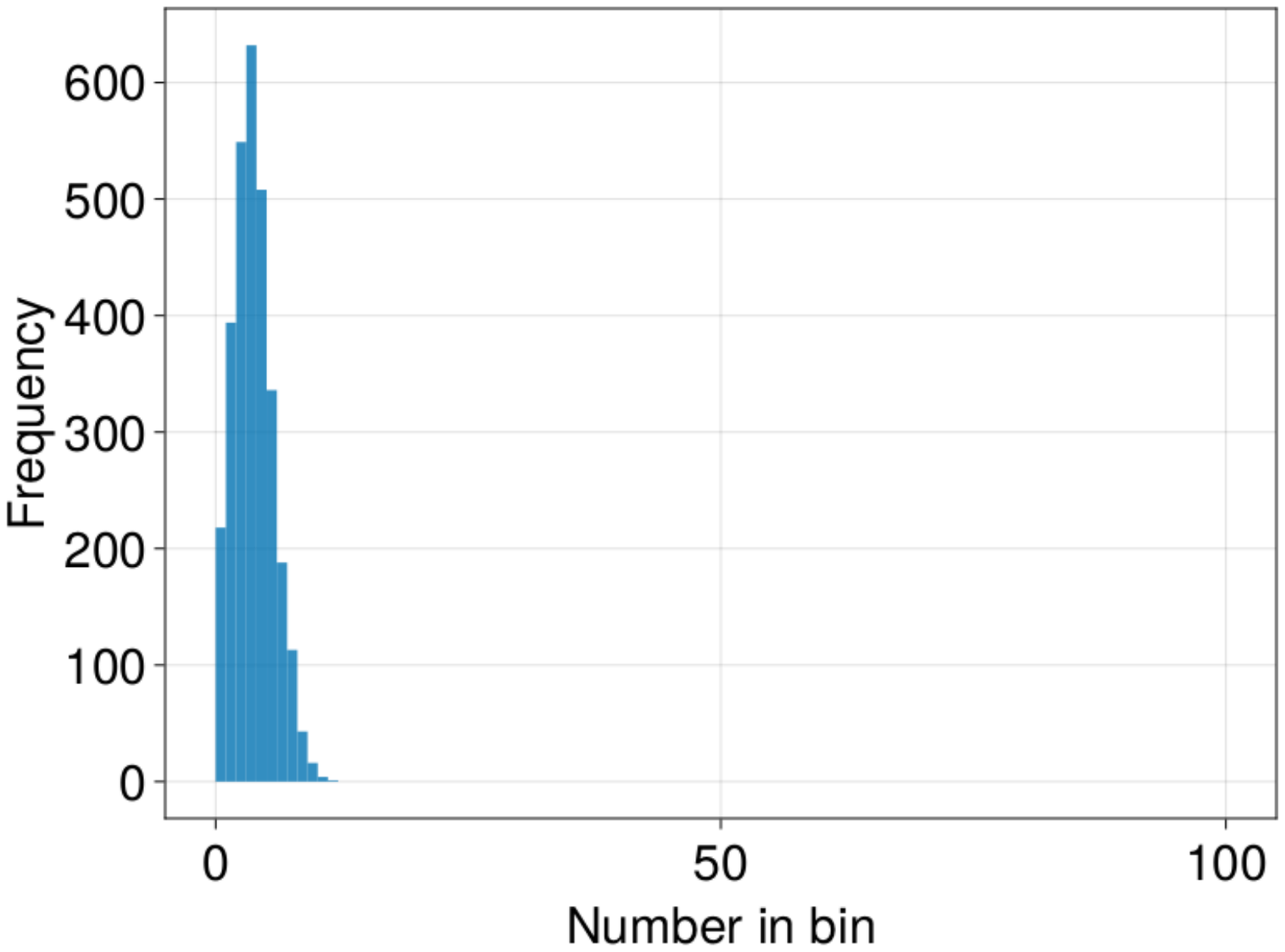}
        \caption{The distribution of the number of agents occupying the unhappiness level of 0.195 to 0.2 for a simulation using a tDOD of $1000\times\sqrt{12}$ over 5000 time steps.}
        \label{fig:single_particle_unhappiness_dist}
    \end{figure}

We therefore posit that the changes in possible unhappiness states for the individual agents characterises the phase changes of the flock - analogous to the way in which the changes in distribution of single particle energy eigenstates indicates changes in the entropy and state of a system in statistical mechanics.

One interesting feature we found through numerical testing was that the turning scheme for stationary agents played a key role in the dynamics and collective behaviour of the system, even when the number of agents that undertook stationary turning per time step was small. In particular, we found that if the agents were programmed to have a bias towards turning in a certain direction, then in the tDOD range of 1800 to 4000, where on average only about 5-10 agents would undertake stationary turning, the resulting flock would exhibit an even stronger form of rotational motion that was uni-directional.
    \begin{figure}[H]
        \centering
        \begin{subfigure}[t]{0.45\textwidth}
            \includegraphics[width=\textwidth]{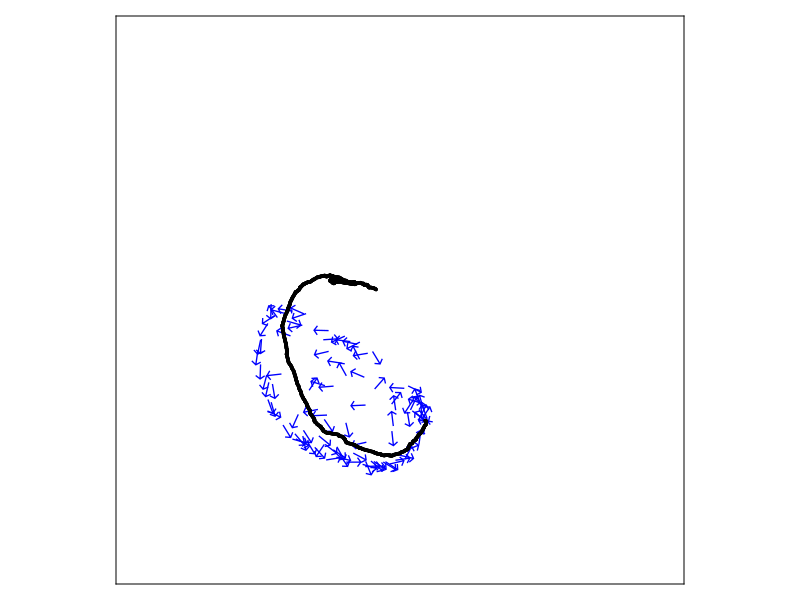}
        \caption{}
        \label{fig:strong_rot_path}
        \end{subfigure}
        \begin{subfigure}[t]{0.45\textwidth}
            \includegraphics[width=\textwidth]{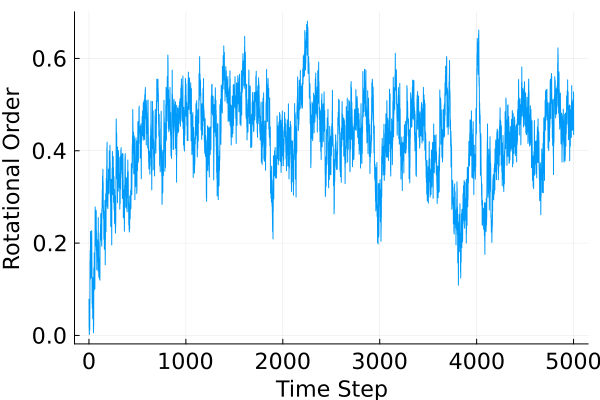}
        \caption{}
        \label{fig:strong_rot_path}
        \end{subfigure}
        \caption{(a) An illustration of the circular trajectory agents can have (shown as the black path) under the strongly rotating movement scheme. The solid black line indicates the path of a single, randomly chosen agent. The domain is of size $800\times800$. (b) An illustration of the evolution of the rotational order for a simulation under the strongly rotating scheme.}
    \end{figure}

A salient question is also what effect the use of a forward bounded DOD has on the dynamics of our flock. Testing the model without forward bounding, we found that  while the mean speed and rotational order of the flock after reaching the equilibrium state is unaffected, the time it takes for the flock to reach the circular expansion phase and non-stationary equilibrium increases. 
Moreover, using non-forward bounded DOD calculations eliminates the pairs flying and milling behaviour of the flock under the biased turning scheme. Even allowing for a more biased turning for stationary agents, the flock would be unable to reproduce milling or strong rotational behaviour with unbounded DOD calculations.

\section{Conclusion}
 In order to drive further collective behaviours after flock formation, we propose a model that generalises the domain minimisation principle underpinning previous SHH models. Agents in the model aim for a specific (not necessarily minimal) domain area, representing a balance between domain minimisation and the desire for agents to maintain inter-agent distances. Moreover, to give a more realistic representation of the field of vision of agents, we bound agent domain calculations to the space in front of them. We do so using an adapted half plane intersection algorithm (given in the appendix) which also allows us to conveniently represent different ranges of vision. Ultimately, we find that the generalised model allows us to overcome the hurdle of flock crystallisation previous SHH models had encountered, with flocks exhibiting motion as a collective. Notably, in a manner similar to phase changes in matter, the collective behaviour of the flock goes through several distinct phases in response to changes in the target DOD. For low target DODs, the flock continues to form a crystallised state previous SHH models found. However, as the target DOD increases, macroscopic movement and circular movement emerges in the flock. Further increases in the tDOD can lead to other effects such as caving-in and pairs flying, before the flock eventually reaches a gaseous phase with movement becoming purely radial. Thus, we find that generalising the domain minimisation principle leads to a much richer range of possible collective behaviours, and offers a platform to explore even more collective behaviours in the future.

\section{CRediT authorship contribution statement}
\textbf{Jesse Zhou}: Wrote the computer code for the half plane intersection algorithm and simulations, performed the numerical simulations, analysed the results, Discussion of the results, Writing - original draft. \textbf{Shannon Dee Algar}: Devised the project, analysed the results, discussion of the results, reviewed and edited original draft. \textbf{Thomas Stemler}: Devised the project, discussion of the results, reviewed and edited original draft.

\section{Competing Interests}
The authors declare no competing interests. 

\section{Acknowledgements}
We received no funding for this project. 

\section{Supplementary Materials}
Attached as videos. 

\newpage
\section{Appendix}
As we see in figure \ref{fig:forward}, our model will utilise the half plane based approach for Voronoi cell calculations to conveniently represent the vision of an agent who can only see ``in front" (i.e, see 90 degrees to either side of the direction they currently face). However, we would like to note that the half plane approach is in fact versatile enough to represent a number of different angles of vision. With humans for example, the field of view is commonly cited to be in the range of $180-220$ degrees \cite{doi:10.1177/2041669520913052}; on the other hand, certain birds have been cited as having an almost complete $360$ degree field of view \cite{MARTIN200825}. If we are willing to overlook the condition of the cell being a Voronoi cell, we can better represent Hamilton's idea of the domain of danger for animals with limited vision by bounding the cell to within the angle of vision of the agent. This can be accommodated by the half plane approach, as it can calculate the area within the bounded domain as the union of multiple convex domains. 

\begin{figure}[H]
        \begin{subfigure}[t]{0.3333\textwidth}
        \centering
            \begin{tikzpicture}
                \draw (5mm,-8.66mm) arc [start angle=-60, end angle=240, radius=10mm];
                \fill[green!] (0,0) -- (5mm, -8.66mm)
            arc [start angle=-60, end angle=240, radius=10mm] -- (0,0);
                \draw (-5mm, -8.66mm) -- (0mm, 0mm);
                \draw (0mm, 0mm) -- (5mm, -8.66mm);
                \filldraw (0,0) circle (1pt)  node[anchor = south east] {$r_i$};
                \draw [->, color = orange, label=left:$H$] (0,0) -- (0mm, 5mm) node[anchor = north west, color = black] {$v_i$};        
    \end{tikzpicture}
        \end{subfigure}
    \begin{subfigure}[t]{0.3333\textwidth}
        \centering
            \begin{tikzpicture}
                \draw (0mm,10mm) arc [start angle= 90, end angle=240, radius=10mm];
                \fill[green!] (0,0) -- (0mm, 10mm)
            arc [start angle=90, end angle=240, radius=10mm] -- (0,0);
                \draw (-5mm, -8.66mm) -- (0mm, 0mm);
                \draw[->,  shorten >= -0.25cm, shorten <=-0.30cm] (0, 0) -- (0mm, 10mm) node[anchor = north west]{$H_{middle}$};
                \draw[->,  shorten >= -0.25cm, shorten <=-0.30cm] (-5mm, -8.66mm) -- (0mm, 0mm) node[anchor = north west]{$H_{left}$};
    \end{tikzpicture}
        \end{subfigure}    
    \begin{subfigure}[t]{0.31\textwidth}
        \centering
            \begin{tikzpicture}
                \draw (0mm,10mm) arc [start angle= 90, end angle=-60, radius=10mm];
                \fill[green!] (0,0) -- (0mm, 10mm)
            arc [start angle=90, end angle=-60, radius=10mm] -- (0,0);
                \draw[->,  shorten >= -0.25cm, shorten <=-0.30cm] (0, 0) -- (0mm, 10mm) node[anchor = north west]{$H_{middle}$};
                \draw[->,  shorten >= -0.25cm, shorten <=-0.30cm] (5mm, -8.66mm) -- (0mm, 0mm) node[anchor = north west]{$H_{right}$};
    \end{tikzpicture}
        \end{subfigure}    
        \caption{(a) The example bounded polygon whose area (highlighted in green) we wish to find. Subfigures (b) and (c) illustrate the decomposition of the bounded area into two convex polygons which can be found through the half plane intersection algorithm. $H_{\mathrm{left}}, H_{\mathrm{right}}, H_{\mathrm{middle}}$ are artificial half planes we use to bound the Voronoi cell calculations to the areas we desire for representing a field of vision.}
        \label{fig:extended_vision}
    \end{figure}
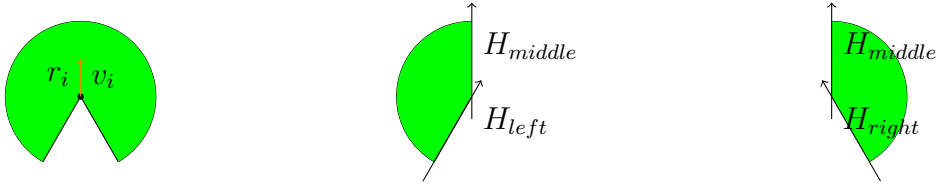

To give an example, suppose that for an agent $i$ we only wanted to find its ``domain of danger" bound within the green area illustrated in the figure \ref{fig:extended_vision}. The half plane approach allows us to calculate this area even though this is not a convex polygon (and therefore not a strict Voronoi cell) by finding the area of two convex shapes split by a half plane

\subsection{The circle bounded half plane intersection method}
\subsubsection{Algorithm terminology}\label{sec:term2}
In order to provide context for the features of our movement model, we first outline the basic definitions and concepts of the half-plane intersection algorithm. For this document, we will denote every half plane as a capital `H' with a greek letter subscript to differentiate different half planes. For example, we can label two half planes as $H_\alpha$ and $H_\beta$. Under the half plane intersection algorithm, every half plane $H_\alpha$ can be defined by a point and a vector (the vector in turn defines the angle of the half plane). We denote the defining point of a given half plane $H_\alpha$ as $\Vec{r}_\alpha$, and the defining vector of the half plane as $\Vec{v}_\alpha$. A point $\Vec{r}$ is said to be \textbf{inside} the half plane $H_\alpha$ if 
    \begin{align}
        \Vec{v}_\alpha \times (\Vec{r} - \Vec{r}_\alpha) \geq 0,
    \end{align}
and conversely, we say that a point $\Vec{r}$ is \textbf{outside} the half plane $H_\alpha$ if
    \begin{align}
        \Vec{v}_{\alpha} \times ((\Vec{r} - \Vec{r}_\alpha) < 0.
    \end{align}

For a given half plane $H_\alpha$ whose ``fence" or defining line has two intersects with a bounding circle associated with each agent's vision, we label one of the intersects as the \textbf{backwards circle intersect}, denoted $\Vec{b}_\alpha$ and the other intersect as the \textbf{forwards circle intersect}, $\Vec{f}_\alpha$. The labelling of the forwards and backwards circle intersects is defined by
    \begin{align}
        \Vec{v}_\alpha \cdot (\Vec{f}_\alpha - \Vec{b}_\alpha) \geq 0.
    \end{align}
In our altered algorithm, for every vertex $\Vec{x}_{\alpha\beta}$ formed between the intersect of the lines defining two half planes $H_{\alpha}, H_{\beta}$, or a half plane line and the bounding circle, we will label one of the half planes to be the \textbf{back half plane}, denoted as $H_b$, and the other half plane to be the \textbf{forward half plane}, denoted by $H_f$. How we choose this labelling will be explained in the algorithm section. We say that a vertex $\Vec{x}_{\alpha\beta}$ is \textbf{valid} if the two criteria are satisfied:
    \begin{align}
        &\Vec{x}_{\alpha\beta} + \Vec{v}_f \in H_b,\\
        &\Vec{x}_{\alpha\beta} - \Vec{v}_b \in H_f.
    \end{align}

 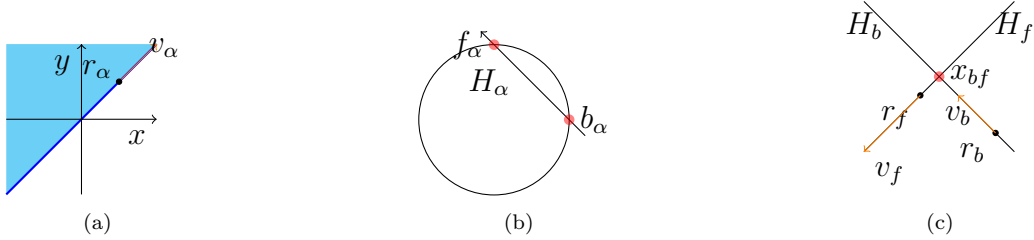
\begin{figure}[H]
        \centering
        \begin{subfigure}[t]{0.3\textwidth}
         \centering
         \begin{tikzpicture}[scale = 1]
\fill[cyan!50]      (-1, -1) -- (1,1) -- (-1,1) -- cycle;
\draw[blue, thick] (-1, -1) -- (1,1) {};
\draw[->] (-1,0) -- (1,0) node[below left] {$x$};
\draw[->] (0,-1) -- (0,1) node[below left] {$y$};
\draw[->, color = orange]  (0.5, 0.5) -- (1,1) ;
\draw (1.1, 0.95) node[color = black]  {$v_\alpha$};
\filldraw (0.5,0.5) circle (1pt) (0.55, 0.40) node[anchor = south east] {$r_\alpha$};
    \end{tikzpicture}
    \caption{}
     \end{subfigure}
     \hfill
     \begin{subfigure}[t]{0.3\textwidth}
         \centering
         \begin{tikzpicture}
          
  \coordinate   (A) at (0,0);
  \coordinate  (B) at (1.0,0.0);
  \coordinate (C) at (1, 0.0);
  \coordinate (D) at (0.0, 1);
  \node (H) [name path=H, draw,circle through=(B)] at (A) {};
  \path [name path=C--D] (C) -- (D);
  \node [label=left:$H_\alpha$] at (0.5, 0.5) {};
  \draw [->, shorten >= -0.25cm, shorten <=-0.30cm, label=left:$H$] (C) -- (D);
  \path [name intersections={of=H and C--D,by={[label=right:$b_\alpha$]I, [label=left:$f_\alpha$]G}}];
  \fill[red,opacity=.5] (G) circle (2pt);
  \fill[red,opacity=.5] (I) circle (2pt);
        \end{tikzpicture}   

        \caption{}
     \end{subfigure}
    \hfill
     \begin{subfigure}[t]{0.3\textwidth}
         \centering
        \begin{tikzpicture}
    \path [name path = Halpha, draw] (1, 1) -- (-1, -1);
    \path [name path = Hbeta, draw] (1, -1) -- (-1, 1);
    \filldraw (0.75, -0.75) circle (1pt) (0.75, -0.75) node[anchor = north east] {$r_b$};
    \draw[->, color = orange]  (0.75, -0.75)  -- (0.25,-0.25) node[anchor = north, color = black] {$v_b$};
    \filldraw (-0.25, -0.25) circle (1pt) (-0.25, -0.25) node[anchor = north east] {$r_f$};
    \draw[->, color = orange]  (-0.25, -0.25)  -- (-1,-1) node[anchor = north west, color = black] {$v_f$};
    \path [name intersections={of=Halpha and Hbeta,by={[label=right:$x_{bf}$]I}}];
     \fill[red,opacity=.5] (I) circle (2pt);
     \node[anchor = north] at (-1, 1) {$H_b$};
     \node [anchor = north] at (1, 1) {$H_f$};
\end{tikzpicture}

         \caption{}
     \end{subfigure}
   \caption{In figure (a), an illustration of how the half plane (coloured blue) is defined with a defining point $\Vec{r}_\alpha$ and vector $\Vec{v}_\alpha$. Figure (b) gives an illustration of the backwards and forwards circle intersects of a half plane $H_\alpha$, denoted by $b_\alpha$ and $f_\alpha$ respectively. (c) Illustration of the labelling of forward and backwards half planes for an intersection $x_{bf}$ between half planes $H_b$ and $H_f$. The forward half plane in this case is $H_f$, while the backwards half plane is $H_b$.}
    \end{figure}
\subsubsection{Algorithm}
We can then calculate the Voronoi cell for a given agent $i$ at a position $\Vec{r}_i$ in the following manner: 
\begin{enumerate}
        \item Find, for all neighbouring agents $j \neq i$ (whose positions we shall denote as $\Vec{r}_j$), the half plane containing the points closer to $i$ than neighbour $j$. Let us label each such half plane by $H_\alpha$. To find the half plane, first calculate the vector from agent $i$ to $j$, $\Vec{r}_{ij} = \Vec{r}_j - \Vec{r}_i$. Then, use $\frac{1}{2}\Vec{r}_{ij}$ as the defining point for the half plane between agents $i$ and $j$. The defining vector can then be calculated as $r_{ij}$ rotated $\pi/2$ radians counter clockwise:
            \begin{align}
                \Vec{v}_\alpha = \begin{bmatrix}
            0 & -1 \\
            1 & 0
        \end{bmatrix} \Vec{r}_{ij}
            \end{align}
        \item Sort the half planes according to the polar angle coordinate of their defining vector. Store this sorted sequence of half planes in a queue called $hq$. 
        \item Create a new queue which will store and keep track of which half planes are relevant to the intersection. Let us denote this queue as $nq$. Also create a queue storing the vertices generated throughout the algorithm. Let us denote this queue as $vq$. In contrast to previous algorithms, the queue for storing the vertices explicitly is needed due to the fact that some vertices of the circle bounded Voronoi cell are not generated as the intersect between half planes - therefore, merely storing the half planes is not sufficient for being able to record the vertices of a circle bounded Voronoi cell.  
        \item Iterate through the front of the $hq$, popping the front as each half plane is processed. Let us denote the front half plane as $H_\alpha$. Before we add $H_\alpha$ to the queue of relevant half planes, we must first remove any previously put down half planes and vertices that will become redundant due to $H_\alpha$. We do so by processing the earliest and latest half planes put down (popping from the front and back of the dequeue)
            \begin{enumerate}
                \item  While the front vertex of $vq$, which we shall denote as $x_\gamma$, is outside of $H_\alpha$, pop the vertex. If the back half plane of $x_\gamma$ is 0, i.e, is the circle, do nothing. If the back half plane of $x_\gamma$ is another half plane from $nq$, then it means that the half plane is redundant, and we can pop that half plane from $nq$. 
                \item While the back vertex of $vq$, which we shall denote as $x_\upsilon$, is outside of $H_\alpha$, pop the vertex from the back of $vq$. If the forwards half plane of $x_\upsilon$ is 0, i.e, is the circle, do nothing. If the forwards half plane of $x_\upsilon$ is another half plane from $nq$, then it means that the half plane is redundant, and we can pop that half plane from the back of $nq$. 
            \end{enumerate}
            
        \item Suppose that the last non-redundant half plane is $H_\beta$. Calculate the intersect of the defining lines of $H_\alpha$ and $H_\beta$. Let us denote their intersection as $x_\epsilon$. Check if the intersection is valid and inside the circle. There are then two posssibilities:
            \begin{enumerate}
                \item If the intersection is valid, add the intersect $x_\epsilon$ to $vq$ as a tuple, with the first element being the intersect coordinates, and the second and third elements being the labels of the backwards and forwards half planes respectively ($x_\epsilon, \beta, \alpha$). Also add the forward circle intersect of $H_\alpha$ to $vq$, with information labelling the backwards and fowards half planes of the vertex as being $0$ and $\alpha$ respectively. Finally, add $H_\alpha$ to $nq$.
                \item However, if the intersection $x_\epsilon$ is not simultaneously valid and inside the circle, check if the backwards circle intersect of $H_\alpha$, $b_\alpha$, is inside $H_\beta$. If so, then add $b_\alpha$ to $vq$ as a tuple $(b_\alpha, \alpha, 0)$ to represent the backwards and forwards half planes of the vertex as being the circle and $H_\alpha$ respectively. Also add the forward circle intersect of $H_\alpha$ to $vq$, with information labelling the backwards and fowards half planes of the vertex as being $0$ and $\alpha$ respectively. Finally, add $H_\alpha$ to $nq$. If $H_\alpha$ does not form a valid intersect inside the circle with $H_\beta$ nor have its backwards circle intersect inside $H_\beta$, then $H_\alpha$ contributes nothing to the intersection, and we can simply pop it off $hq$ with no further action. 
            \end{enumerate}
        If $H_\beta$ does not exist, i.e, $H_\alpha$ has made all previously relevant half planes redundant, then we simply add $H_\alpha$ and its circle intersects to queues $nq, vq$ respectively. So let the backwards and forwards circle intersects of $H_\alpha$ be denoted by $b_1, f_1$. To $vq$, add these vertices as tuples with information to represent the backwards and forwards half planes of each vertex as $(b_1, 0, 1)$ and $(f_1, 1, 0)$. So the tuple $(b_1, 0, 1)$ represents the fact that the vertex $b_1$ has backwards half plane of $0$, i.e, the circle, and forwards half plane of $H_1$. 
        \item Repeat steps 5 and 6 until all half planes from $hq$ have been processed. 
        \item Do a final clean up from the front: using the half plane at the front of the queue $nq$, which we denote $H_\mathrm{front}$, while the vertices at the back of $vq$ are outside $H_\mathrm{front}$, pop the vertex, and if the forwards half plane of the vertex is non-zero, pop the back of $nq$ as well, since this corresponds to one of the half planes at the back of $nq$ being made redundant.  
        \item Do a final clean up from the back: using the half plane at the back of the queue $nq$, which we denote $H_\mathrm{back}$, while the vertices at the front of $vq$ are outside $H_\mathrm{back}$, pop the vertex, and if the backwards half plane of the vertex is non-zero, pop the front of $nq$ as well, since this corresponds to one of the half planes at the front of $nq$ being made redundant.
        \item Finally, add the intersection between the front and back half planes of $nq$ if it is a valid intersect and inside the circle. Otherwise, do nothing.  
    \end{enumerate}
Since each half plane can at most contribute two vertices, and each vertex can only be added or popped once, the bottleneck in our algorithm occurs in sorting the half planes. Thus, the overall complexity of our algorithm is $O(n\log(n))$. 

\subsection{Algorithm Illustration}
Here, we give an example run-through of how our algorithm calculates the circle and forward bounded Voronoi cell of an agent. In this example, we show the algorithm calculating the cell of agent 1 given the agent positions as shown in figure \ref{fig:cell_illustration_0}.
    \begin{figure}[H]
        \begin{subfigure}[t]{0.45\textwidth}
            \includegraphics[width=\textwidth]{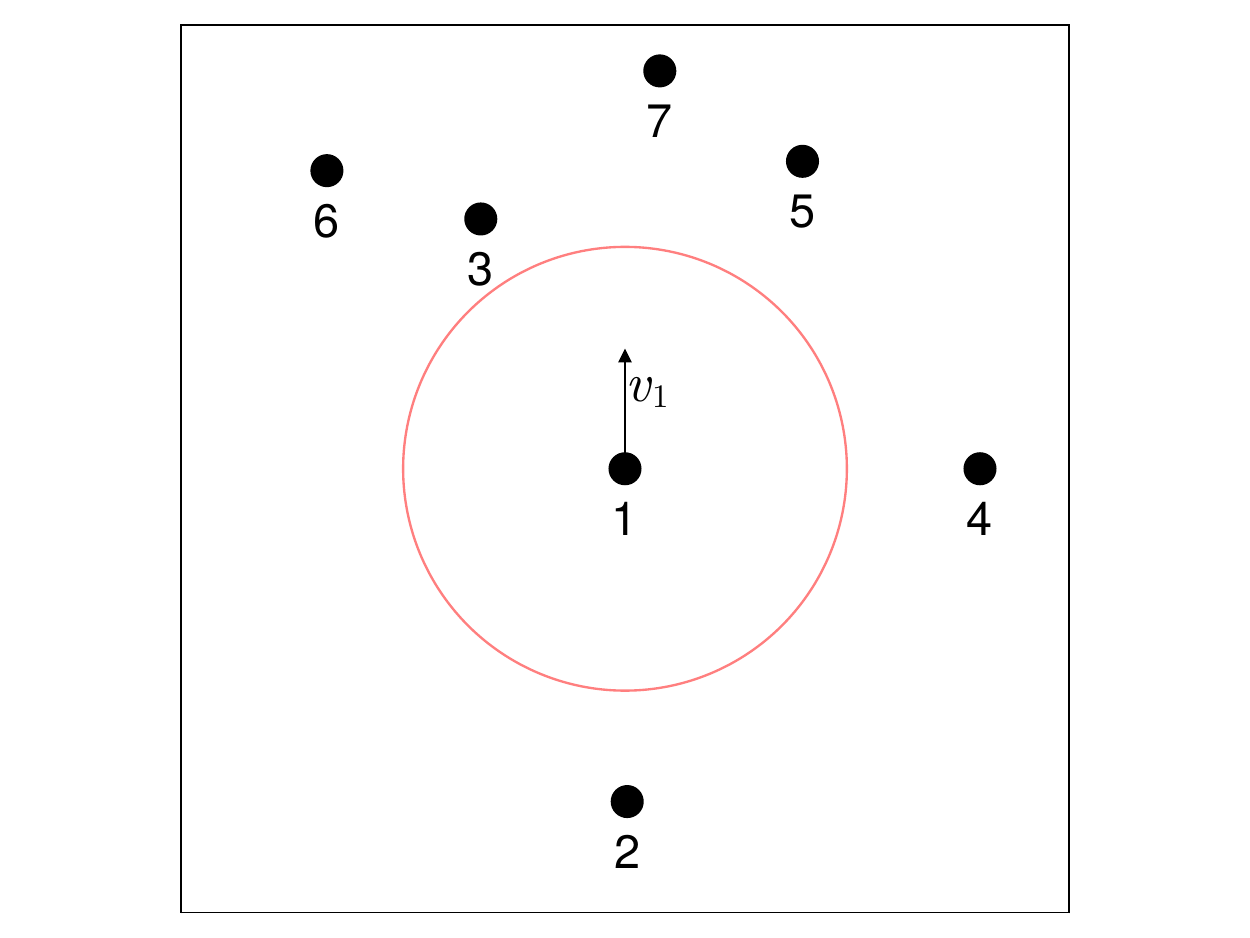}
        \caption{An illustration of our example flock model. We will consider the construction of the Voronoi cell for agent 1, with a velocity of $v_1$ as indicated in the figure. The agent positions are shown as the black dots, while the semi-transparent red circle represents the circle of vision of agent 1. }
        \label{fig:cell_illustration_0}
        \end{subfigure}
        \hfill
        \begin{subfigure}[t]{0.45\textwidth}
            \includegraphics[width=\textwidth]{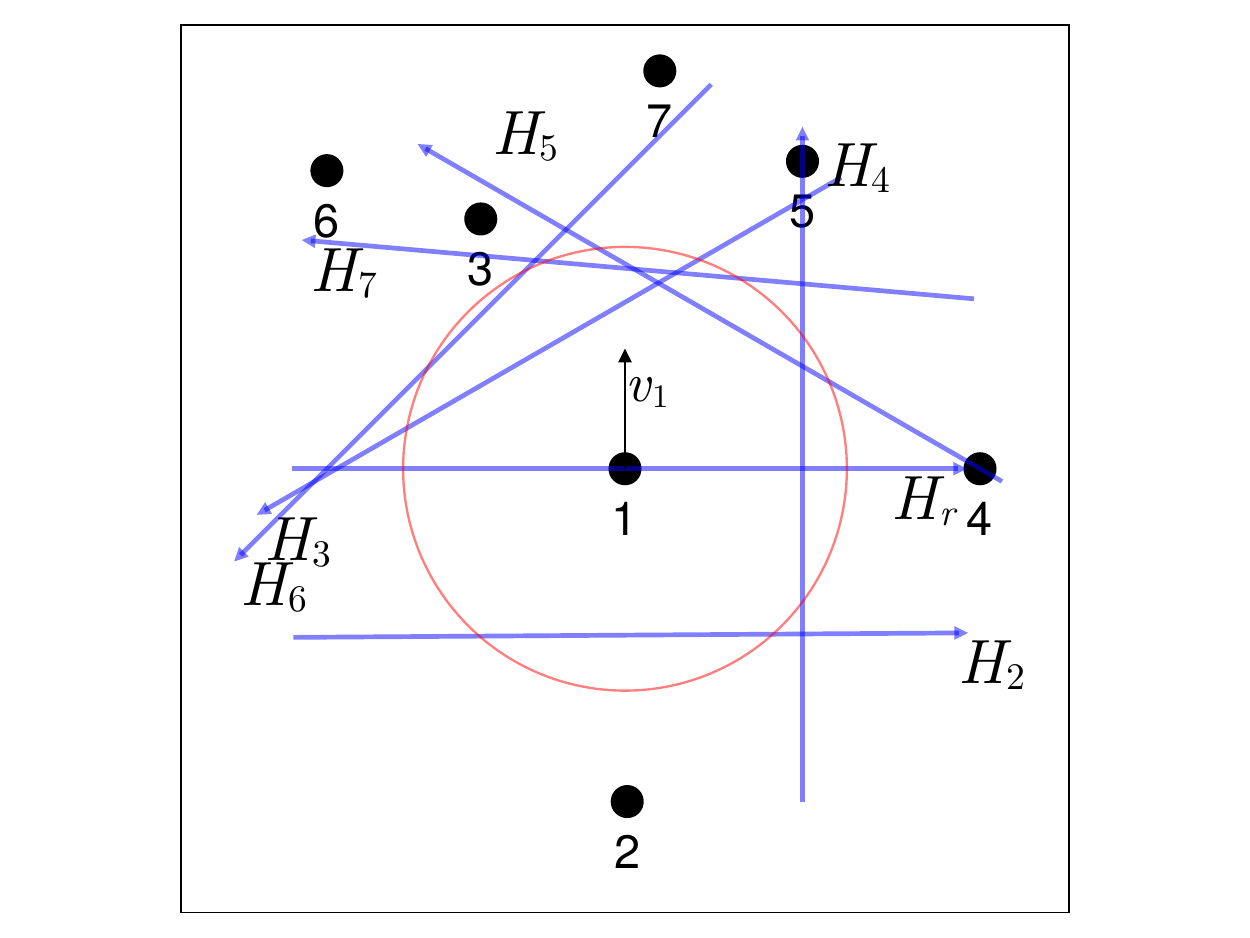}
        \caption{An illustration of the half planes generated between agent one and each of its neighbouring agents, represented by an arrow. The half plane associated with neighbour $j$ is labeled as $H_j$. Note that we have represented the artificial or ``relic" half plane used to forward bound the Voronoi cell as $H_r$.}
        \label{fig:cell_illustration_0_hp}
        \end{subfigure}
    \end{figure}
    
    \begin{figure}[H]
        \ContinuedFloat
        \begin{subfigure}[t]{0.45\textwidth}
            \includegraphics[width=\textwidth]{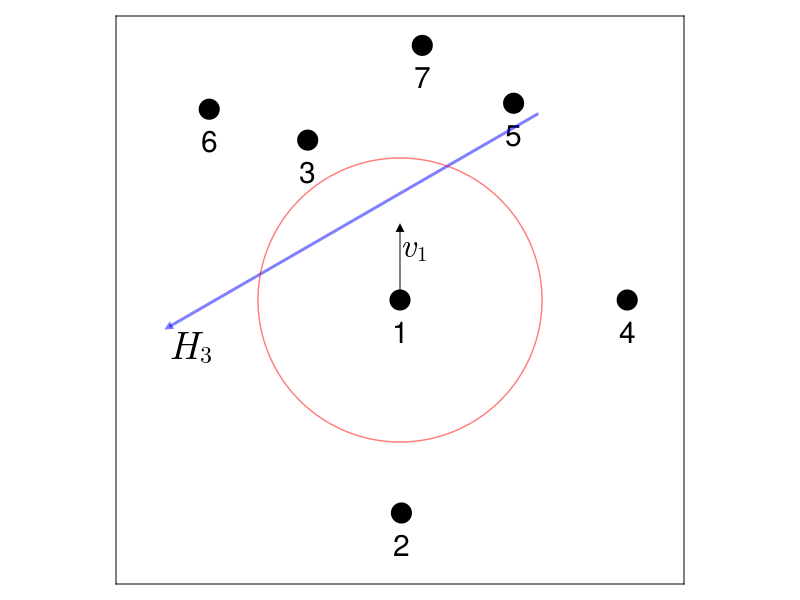}
        \caption{We will consider each half plane sorted according to the angle of their defining vector (indicated by the direction of the arrow). We start off with half plane $H_3$. There are no previously put down vertices or half planes, and so we automatically add $H_3$ and its backward and forwards circle intersects.}
        \label{fig:cell_illustration_1}
        \end{subfigure}
        \hfill
        \begin{subfigure}[t]{0.45\textwidth}
            \includegraphics[width=\textwidth]{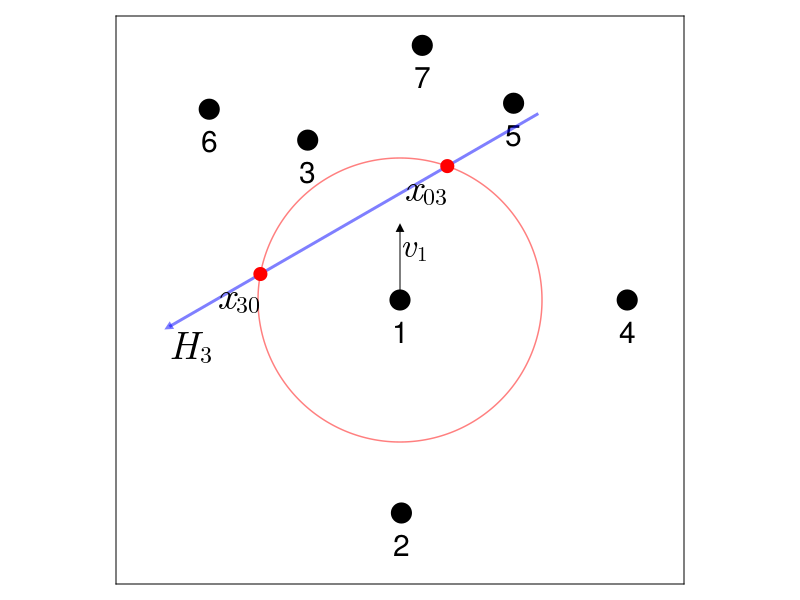}
        \caption{An illustration of the relevant half planes and vertices after considering $H_3$, which are its circle intersects, $x_{03}$ and $x_{30}$.}
        \label{fig:cell_illustration_1_post}
        \end{subfigure}
    \end{figure}
    
    \begin{figure}[H]
    \ContinuedFloat
        \begin{subfigure}[t]{0.45\textwidth}
            \includegraphics[width=\textwidth]{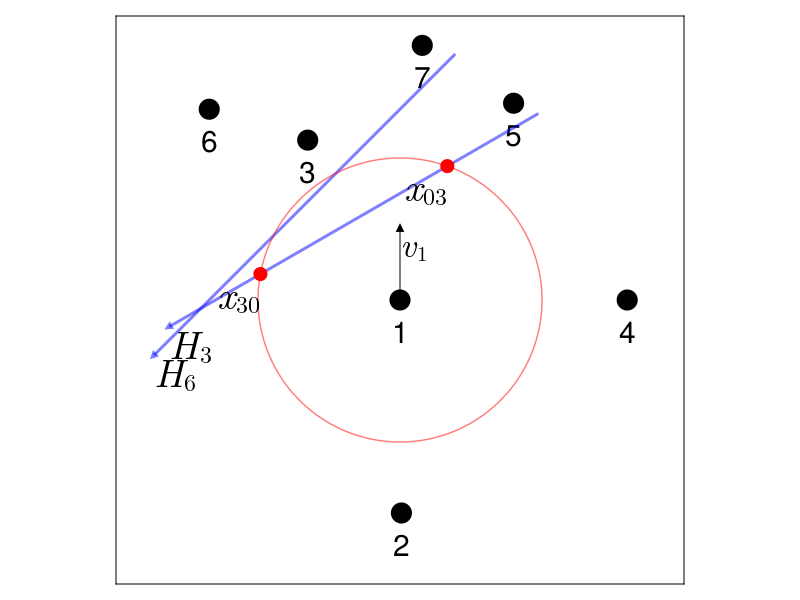}
        \caption{We then move onto half plane $H_6$, shown in figure \ref{fig:cell_illustration_2}. We can see that it makes none of the previously added vertices redundant. However, it is also apparent that its backwards circle intersect lies outside of $H_3$, which therefore tells us that $H_6$ is redundant.}
        \label{fig:cell_illustration_2}
        \end{subfigure}
        \hfill
        \begin{subfigure}[t]{0.45\textwidth}
            \includegraphics[width=\textwidth]{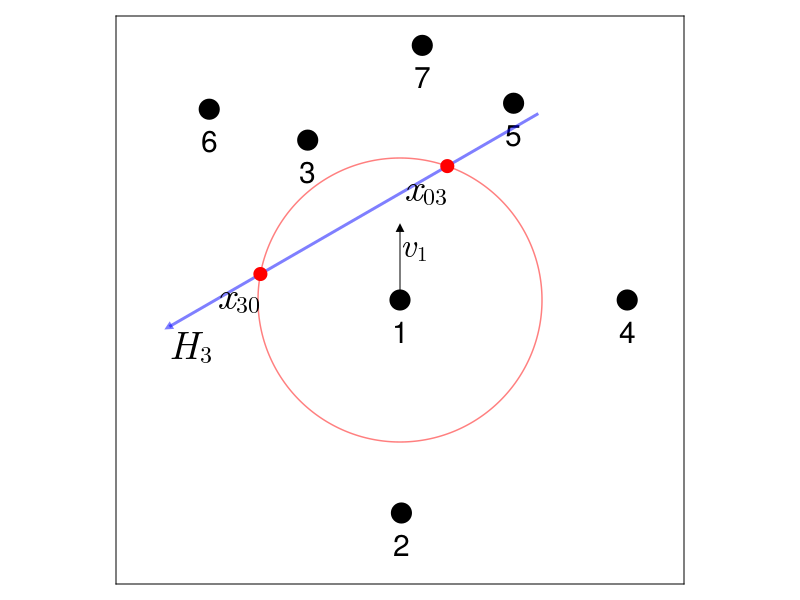}
        \caption{The relevant half planes and vertices is therefore unchanged from the previous step.}
        \label{fig:cell_illustration_2_post}
        \end{subfigure}
        
        \end{figure}

        \begin{figure}[H]
        \ContinuedFloat
        \begin{subfigure}[t]{0.45\textwidth}
            \includegraphics[width=\textwidth]{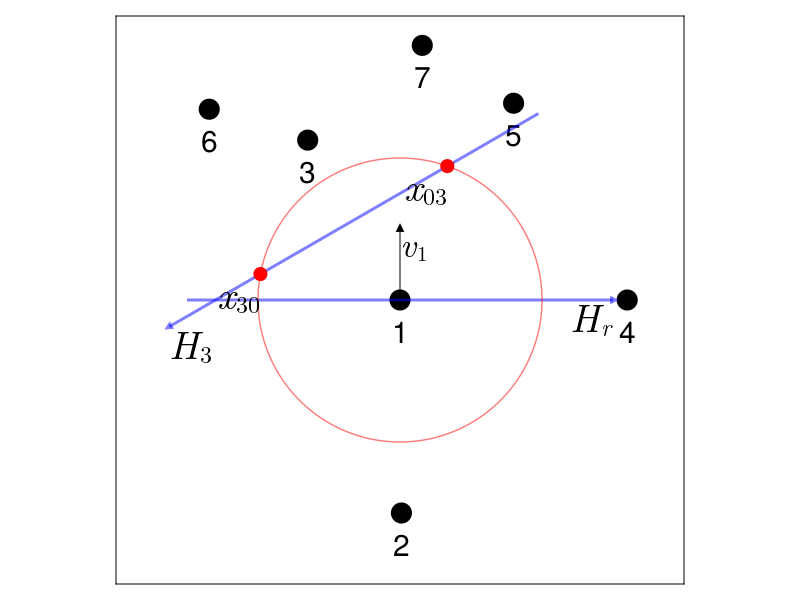}
        \caption{The next half plane to consider is $H_r$, which represents the artificial half plane used to bound our Voronoi cell calculation to space in front of agent 1. We can see that it makes no previously added vertices redundant. We note that it does not have a valid intersect inside the bounding circle with previous half plane $H_3$, but it's backwards circle intersect is still within $H_3$. This means that unlike $H_6$, $H_r$ is not made redundant by $H_3$, and so we can add its backwards and forwards circle intersects to the cell.}
        \label{fig:cell_illustration_3}
        \end{subfigure}
        \hfill
        \begin{subfigure}[t]{0.45\textwidth}
            \includegraphics[width=\textwidth]{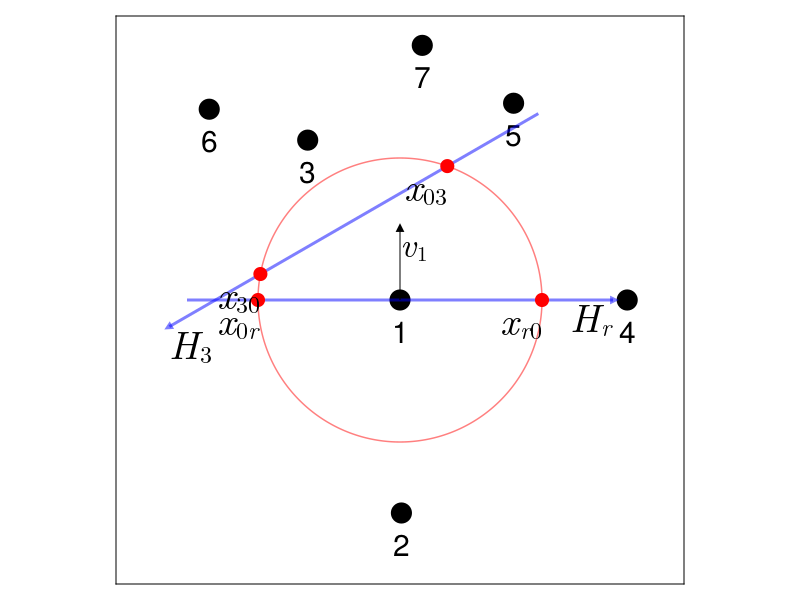}
        \caption{Illustration of the relevant half planes and vertices after considering $H_r$. We add the backwards and forwards circle intersects of $H_r$, $x_{0r}$ and $x_{r0}$ respectively.}
        \label{fig:cell_illustration_3_post}
        \end{subfigure}
    \end{figure}
        
    \begin{figure}[H]
    \ContinuedFloat
        \begin{subfigure}[t]{0.45\textwidth}
            \includegraphics[width=\textwidth]{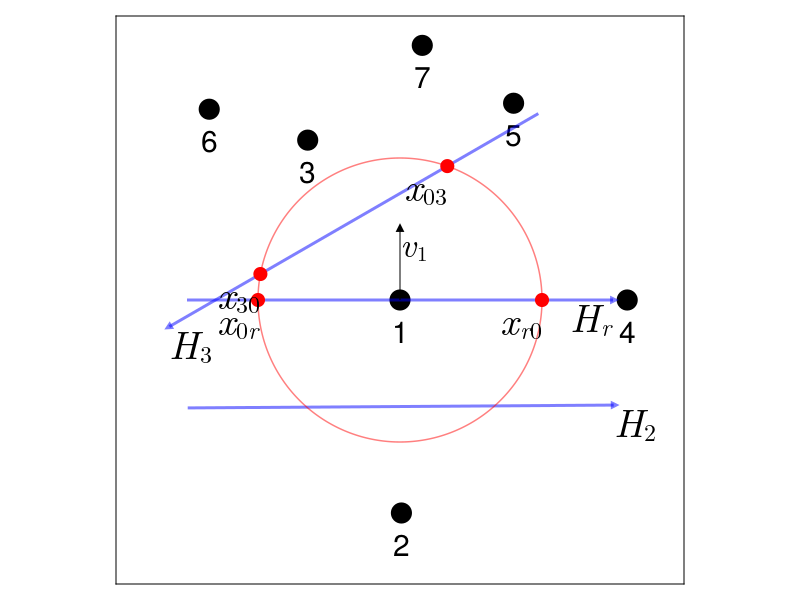}
        \caption{We then consider half plane $H_2$. $H_2$, like $H_6$, is made redundant by $H_r$.} 
        \label{fig:cell_illustration_4}
        \end{subfigure}
        \hfill
        \begin{subfigure}[t]{0.45\textwidth}
            \includegraphics[width=\textwidth]{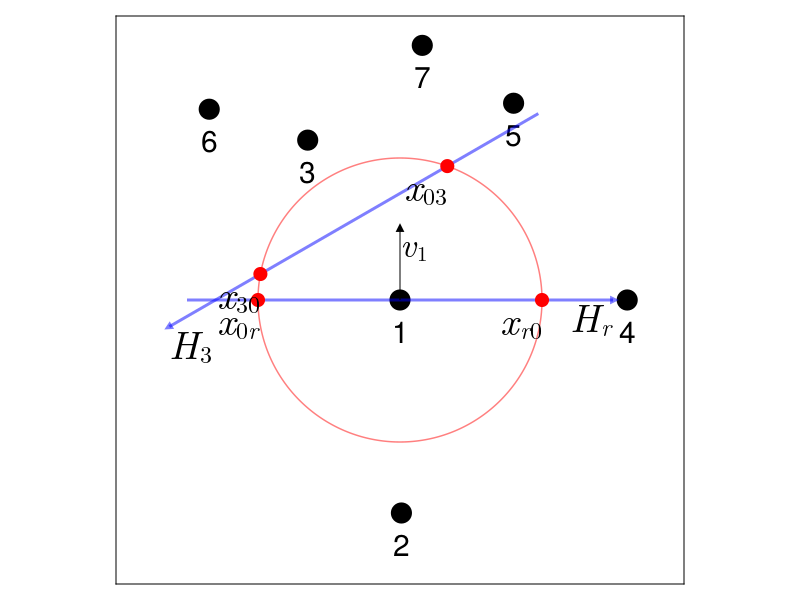}
        \caption{Therefore, the cell after considering  $H_2$ has no changes with the previous step}
        \label{fig:cell_illustration_4_post}
        \end{subfigure}
    \end{figure}
    
    \begin{figure}[H]
    \ContinuedFloat
        \begin{subfigure}[t]{0.45\textwidth}
            \includegraphics[width=\textwidth]{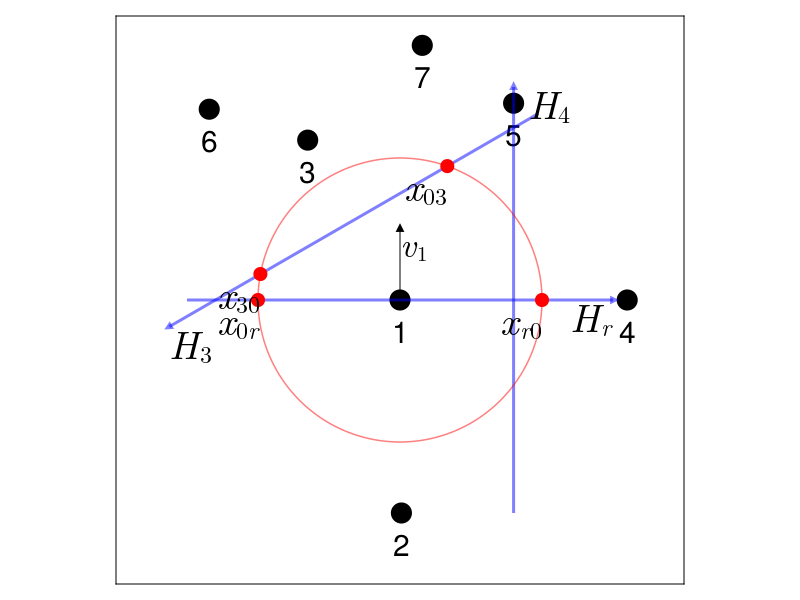}
        \caption{We then consider the next half plane, $H_4$, and we see that it makes redundant vertex $x_{\mathrm{r0}}$. However, since the forward half plane of the vertex is simply the circle, $H_r$ is not made redundant. Moreover, in this case, $H_4$ has a valid intersect with $H_r$ inside the circle.}
        \label{fig:cell_illustration_5}
        \end{subfigure}
        \hfill
        \begin{subfigure}[t]{0.45\textwidth}
            \includegraphics[width=\textwidth]{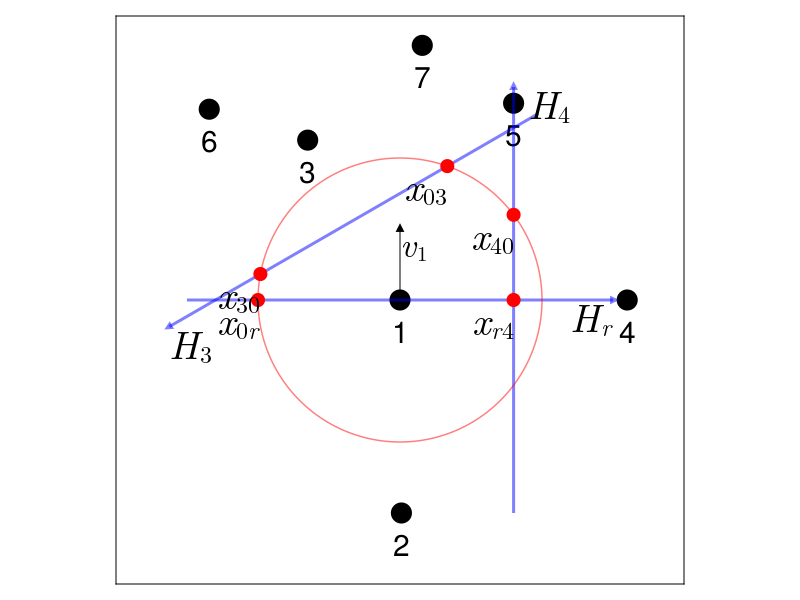}
        \caption{Therefore, we add the intersect $x_{\mathrm{r4}}$ as well as the forward circle intersect of $H_4$.}
        \label{fig:cell_illustration_5_post}
        \end{subfigure}
    \end{figure}

    \begin{figure}[H]
        \ContinuedFloat
        \begin{subfigure}[t]{0.45\textwidth}
            \includegraphics[width=\textwidth]{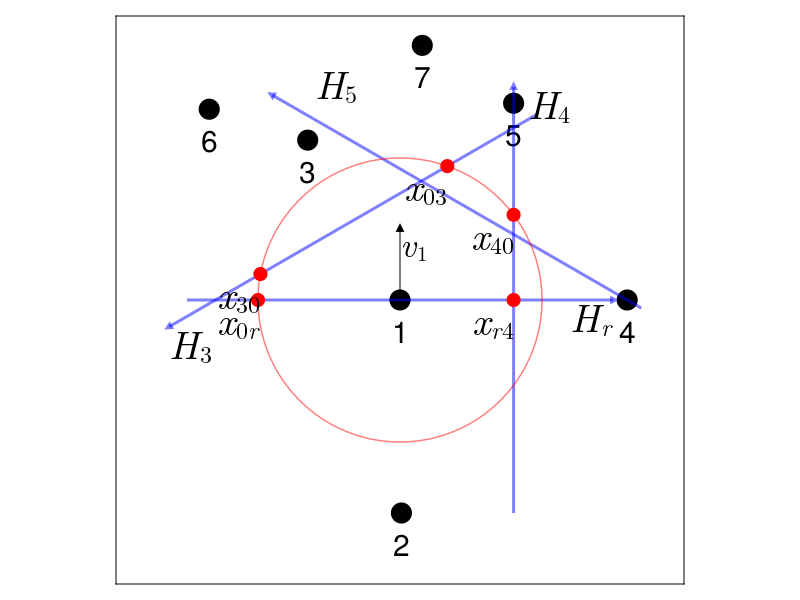}
        \caption{We then consider half plane $H_5$, which makes redundant $x_{03}$ and $x_{40}$}
        \label{fig:cell_illustration_6}
        \end{subfigure}
        \hfill
        \begin{subfigure}[t]{0.45\textwidth}
            \includegraphics[width=\textwidth]{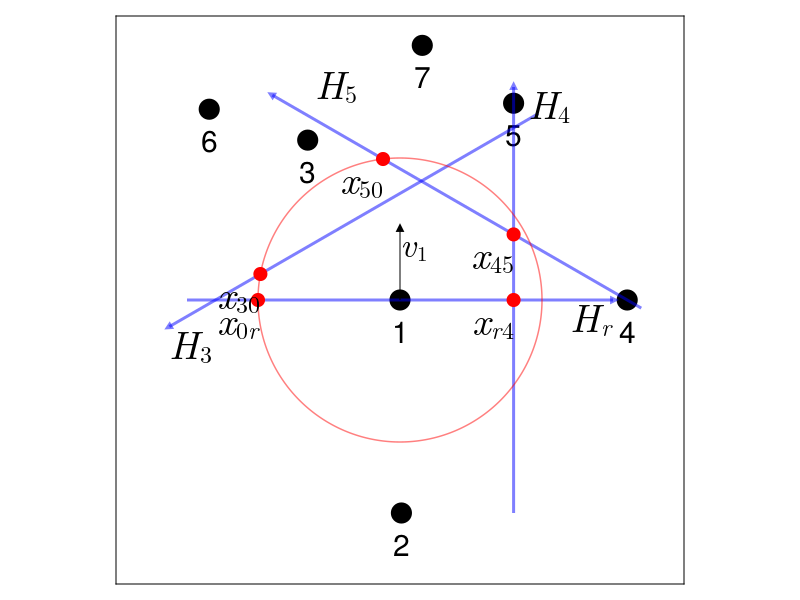}
        \caption{We then add its intersect with $H_4$ as well as its forwards circle intersect, $x_{50}$.}
        \label{fig:cell_illustration_6_post}
        \end{subfigure}
    \end{figure}
        
    \begin{figure}[H]
    \ContinuedFloat
        \begin{subfigure}[t]{0.45\textwidth}
            \includegraphics[width=\textwidth]{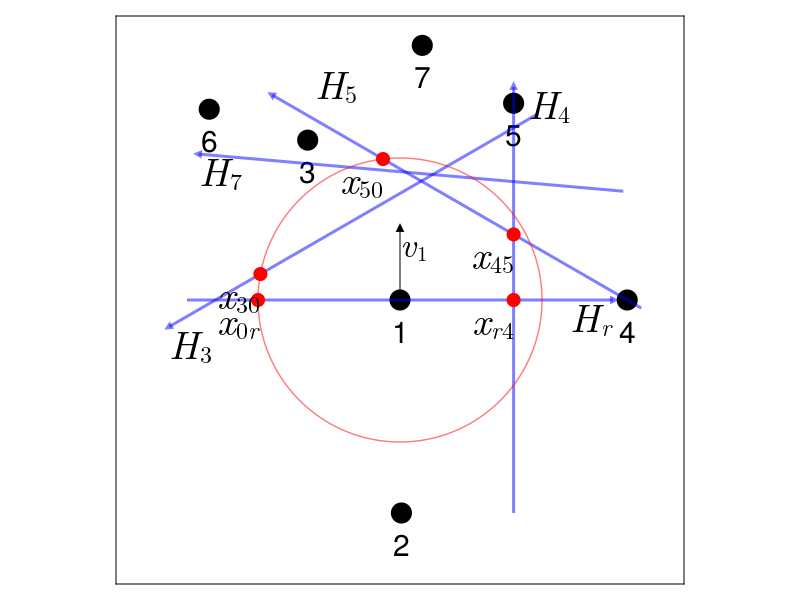}
        \caption{The final half plane we consider is $H_7$, which makes redundant $x_{50}$. We note that our algorithm detects that $H_7$ does have a valid intersect inside the circle with $H_5$.}
        \label{fig:cell_illustration_7}
        \end{subfigure}
        \hfill
        \begin{subfigure}[t]{0.45\textwidth}
            \includegraphics[width=\textwidth]{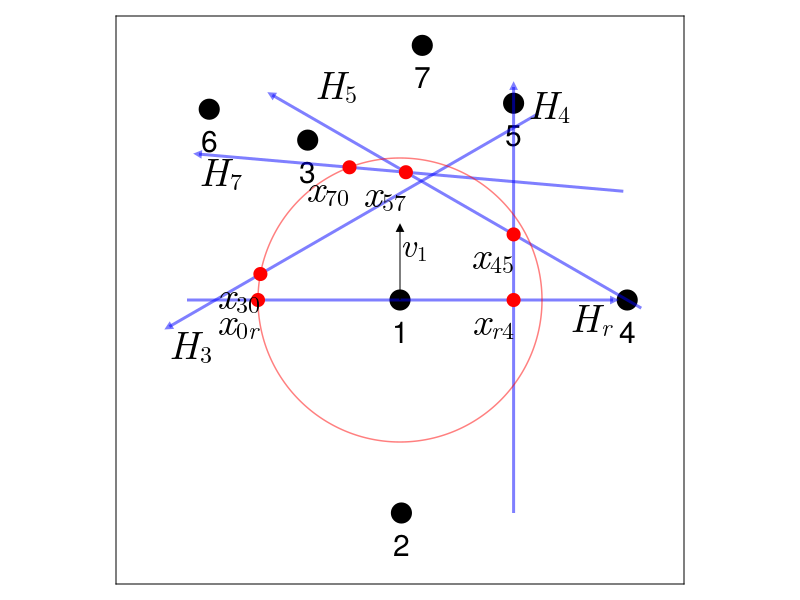}
        \caption{Therefore, we can add the intersect between $H_5$ and $H_7$ along with the forwards circle intersect of $H_7$ to the dequeues of relevant vertices.}
        \label{fig:cell_illustration_7_post}
        \end{subfigure}
    \end{figure}
        
    \begin{figure}[H]
    \ContinuedFloat
        \begin{subfigure}[t]{0.45\textwidth}
            \includegraphics[width=\textwidth]{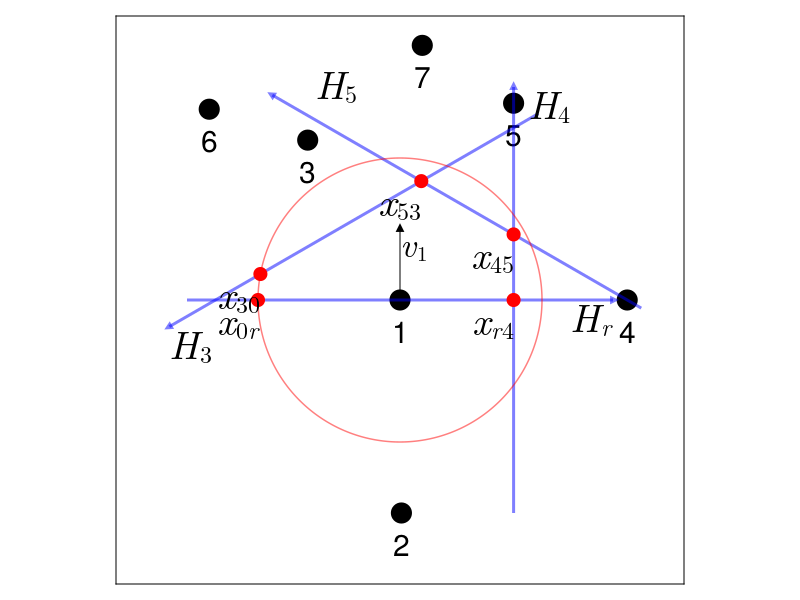}
        \caption{However, from \ref{fig:cell_illustration_7_post}, it is evident that $H_7$ is in fact redundant by a combination of $H_3$ and $H_5$. This illustrates the need for the final cleanup phase of the algorithm, as we use $H_3$ to make $x_{70}$ and $x_{57}$ redundant, and therefore also remove $H_7$. Finally, we calculate the intersect between the earliest and latest relevant half planes, $H_3$ and $H_5$, adding the final vertex $x_{53}$.}
        \label{fig:cell_illustration_final}
        \end{subfigure}
        \hfill
        \begin{subfigure}[t]{0.45\textwidth}
            \includegraphics[width=\textwidth]{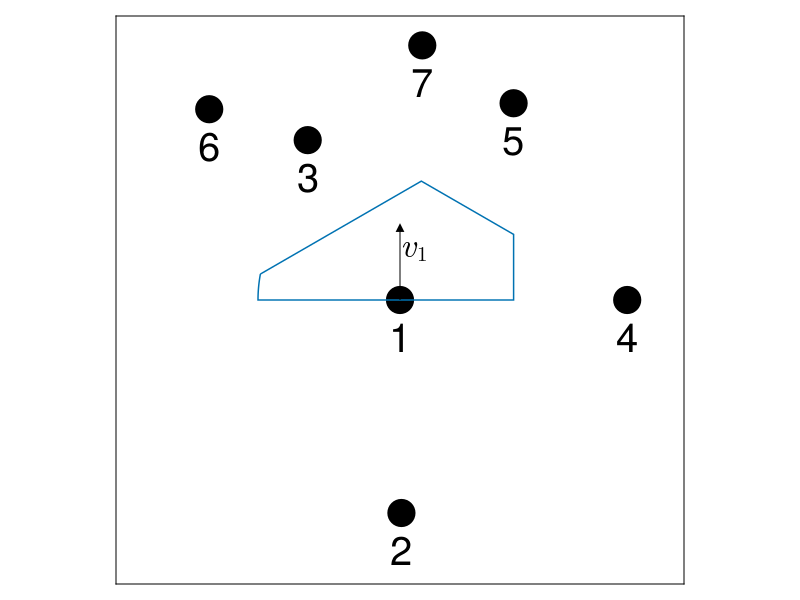}
        \caption{An illustration of our final cell.}
        \label{fig:cell_final}
        \end{subfigure}
    \end{figure}
    
\newpage
\bibliographystyle{unsrtnat}
\bibliography{refs}

\end{document}